# Assessing Engraftment Following Fecal Microbiota Transplant

Chloe Herman[1,2], Bridget M. Barker[1], Thais F. Bartelli[3], Vidhi Chandra[3], Rosa Krajmalnik-Brown[4,5], Mary Jewell[6], Le Li[3], Chen Liao[7], Florencia McAllister[3,8,9], Khemlal Nirmalkar[4], Joao B. Xavier[7], J. Gregory Caporaso[1,2,10]

1. Pathogen and Microbiome Institute, Northern Arizona University, Flagstaff, AZ, USA.
2. School of Informatics, Computing and Cyber Systems, Northern Arizona University, Flagstaff, AZ, USA
3. Department of Clinical Cancer Prevention, University of Texas MD Anderson Cancer Center, Houston, TX, USA.
4. Biodesign Center for Health Through Microbiomes, Arizona State University, Tempe, AZ, U.S.A.
5. School of Sustainable Engineering and the Built Environment, Arizona State University, Tempe, AZ, U.S.A.
6. Independent author, Salt Lake City, UT, USA.
7. Program for Computational and Systems Biology, Memorial Sloan-Kettering Cancer Center, New York, NY, USA.
8. Department of Gastrointestinal Medical Oncology, University of Texas MD Anderson Cancer Center, Houston, TX, USA.
9. Department of Immunology, The University of Texas MD Anderson Cancer Center, Houston, TX, USA.
10. Department of Biological Sciences, Northern Arizona University, Flagstaff, AZ, USA.
11. Corresponding author: greg.caporaso@nau.edu

// Abstract

Fecal Microbiota Transplant (FMT) is an FDA approved treatment for recurrent *Clostridium difficile* infections, and is being explored for other clinical applications, from alleviating digestive and neurological disorders, to priming the microbiome for cancer treatment, and restoring microbiomes impacted by cancer treatment.

Quantifying the extent of engraftment following an FMT is important in determining if a recipient didn't respond because the engrafted microbiome didn't produce the desired outcomes (a successful FMT, but negative treatment outcome), or the microbiome didn't engraft (an unsuccessful FMT and negative treatment outcome). The lack of a consistent methodology for quantifying FMT engraftment extent hinders the assessment of FMT success and its relation to clinical outcomes, and presents challenges for comparing FMT results and protocols across studies.

Here we review 46 studies of FMT in humans and model organisms and group their approaches for assessing the extent to which an FMT engrafts into three criteria: 1) *Chimeric Asymmetric Community Coalescence* investigates microbiome shifts following FMT engraftment using methods such as alpha diversity comparisons, beta diversity comparisons, and microbiome source tracking. 2) *Donated Microbiome Indicator Features* tracks donated microbiome features (e.g., amplicon sequence variants or species of interest) as a signal of engraftment with methods such as differential abundance testing based on the current sample collection, or tracking changes in feature abundances that have been previously identified (e.g., from FMT or disease-relevant literature). 3) *Temporal Stability* examines how resistant post-FMT recipient's microbiomes are to reverting back to their baseline microbiome. Individually, these criteria each highlight a critical aspect of microbiome engraftment; investigated together, however, they provide a clearer assessment of microbiome engraftment.

We discuss the pros and cons of each of these criteria, providing illustrative examples of their application. We also introduce key terminology and recommendations on how FMT studies can be analyzed for rigorous engraftment extent assessment.

# Introduction

The human microbiome is a complex ecosystem composed of microorganisms that can be beneficial, neutral, or detrimental to their host. The gut microbiome, the most populous microbial community living in human organisms, impacts diverse aspects of our health, from host immune response[1,2]; inflammatory bowel diseases like Crohn's Disease and ulcerative colitis[3,4]; metabolic diseases like obesity and diabetes[3,5];autoimmune diseases[6–8]; diverse cancers[9–12]; and even neurological conditions through the gut-brain axis.[13–17]

The many links between the gut microbiome and human health have raised the promise of altering the microbiome to improve health[18,19]. FMTs attempt to replace a recipient's gut microbiome with that of a "donor" by transplanting microbes derived from fecal material (or fecal material as a whole) from a donor to a recipient through endoscopies or oral capsule.[20] The donor, in this case, could be a "healthy" individual, an individual who responded to treatment, or any individual (including the same individual at an earlier time) whose microbiome researchers want to recreate. Before donation, the donated microbiome is typically screened for specific bacterial pathogens, viruses, parasites, and general health metrics.[20]

The USA Food and Drug Administration (FDA) approved FMTs in 2013 to treat recurrent *Clostridium difficile* infection when other treatments (such as antibiotics) failed[21] because of striking findings of its efficacy under those circumstances. (Gut microbiome compositions that have detrimental effects on host health, such as those dominated by *C. difficile*, are often referred to as dysbioses[22], though "dysbiosis" is not a specifically defined term or condition, and we will avoid it in this work[23].) There were significant obstacles to implementing FMTs at the time, including the safety of the donated microbiome, the material preparation for transplantation, deciding the most suitable administration route, and other aspects of the treatment protocol.[20] More recently, in 2022, the first FMT microbiota product was approved by the FDA for treating *C. difficile* infection[24], and in 2023, the first oral FMT microbiota product was approved.[25] In contrast to the 2013 approval, this treatment uses stool from pre-approved donated microbiomes and is rectally administered, making FMTs more accessible and standardized.[24] In addition to treating *C. difficile* infections, FMT had preliminary successes in improving Immune Checkpoint Inhibitors (ICPI) responses[26–28], remediation of tumor growth[29], treating irritable bowel syndrome[27,30], and improving aging hallmarks[31], diabetes[32], behavioral and digestive symptoms associated with autism spectrum disorder[33,34], Alzheimer's disease[35], re-establishing gut microbiota after allogeneic hematopoietic cell transplant[36–39], and more.

One of the most exciting applications of FMT is in cancer treatment, including improving the efficacy of immunotherapy, easing the recovery from cancer treatment, and slowing the pace of tumor growth. In examples of prepping the microbiome for cancer therapies, melanoma patients who have become refractory to anti-PD1 therapy have responded after receiving an FMT from a patient who had previously responded to anti-PD1 therapy [12,26] Other teams have used FMTs to re-establish the microbiome after cancer therapies[36–40]. For example, Defilipp et al.[36] administered an FMT after Allogeneic Hematopoietic Stem Cell Transplant (allo-HCT), an extremely invasive treatment involving chemotherapy and stem cell transplantation, which can be curative for a variety of blood cancers. The team found that FMT intervention helped increase microbiome diversity following the allo-HCT, and was associated with an increased success rate of the treatment. FMT intervention has also been observed to alleviate graft-versus-host disease symptoms after allo-HCT.[37] Finally, FMT may also stem tumor growth. Riquelme et al. (2019)[29] demonstrated that transplants from short-term pancreatic ductal adenocarcinoma (PDAC) survivors correlated with increased tumor size in mouse models, while transplants from long-term PDAC survivors correlated with decreased tumor size and increased immunoactivation. We also hypothesize that FMT or other microbiome alteration techniques could be useful in cancer prevention[41,42], which is generally considered to be more effective at saving lives than cancer treatment[43], though considerable research on personalized optimizations of gut microbiome health is still needed to advance this goal.

# Assessing FMT Success

In an FMT study, the success of microbiome engraftment is often conflated with the success in improving the target clinical outcomes: FMT is considered successful when the desired clinical outcome is achieved. But in order to study the factors driving FMT success, we should evaluate microbiome engraftment independently of the target clinical outcome. A negative target outcome could occur despite FMT success or may indicate a failed FMT engraftment (Figure 1). Still, we lack a standard approach to quantify FMT engraftment extent.

Here, we describe approaches previously used by different FMT studies to assess engraftment extent. We identify three criteria commonly used to quantify engraftment extent, though most studies use only one or two of these criteria. We conclude by offering recommendations for future studies assessing microbiome engraftment, beginning with factors that should be considered during experimental design.

## Terminology: Species, Sequence Variants, and Features

The units of observation in microbiome studies, and the terminology used by researchers, vary with the technology used to profile the microbiome. In small subunit ribosomal RNA amplicon (SSU rRNA, or 16S rRNA) studies, the currently preferred unit of study is the amplicon sequence variant[44], or ASV—a unique sequence determined after data quality control. Before this approach, SSU rRNA studies grouped sequence variants into operational taxonomic units (OTUs), generally by clustering sequences at some percent identity; this approach is out of favor because it reduces the taxonomic resolution unnecessarily. In a microbiome metagenomics survey, sequence reads, assembled contigs, or metagenome-assembled genomes (MAGs), are generally assigned taxonomy, and a taxonomic unit such as genus, species, or strain is used as the unit of observation. Alternatively, in a microbiome metagenomics or metatranscriptomics survey, functional characteristics of a microbiome, such as observed or active genes or pathways, could be the units of observation. In a mass-spectrometry-based metaproteomics or metabolomics study, individual peptides, proteins, or small molecules could be the units of observation.

Despite the difference across these units, many of the downstream analyses are very similar. Here we borrow, the intentionally general term, *feature* from the machine learning field to refer to any of these units. The term *feature table* will be used to describe a table of feature frequencies (counts), relative abundances, or presence/absence information on a per sample basis, for all samples in a given study.

## Clinical Response

If a treatment achieves the desired outcome, we define that as **clinical response**, while **clinical non-response** is defined by the treatment not achieving the desired response. FMT studies often track a patient's response to the FMT by tracking disease status, although this doesn't address whether the FMT treatment actually resulted in the transplant and colonization of a microbiome. A **clinical non-responder** (someone who received the FMT but had no improvement in disease status or symptoms), could have not responded because FMT is not an effective treatment for their condition, or because the FMT itself was unsuccessful, in which case we are not able to assess whether FMT is an effective treatment for their condition. For example, Khanna et al. (2022)[45] considered a successful clinical outcome to be no occurrences of a *C. difficile* infection for six months after the FMT intervention. Over a quarter of participants in their study were clinical non-responders, but without knowing whether or not the microbiome successfully transplanted, we don't know if it's worth subjecting these participants to the treatment again (perhaps with alterations to the administration protocol, such as switching from oral to rectal administration, or modifying the dose), or whether other options should be considered.

## Criteria for assessing microbiome engraftment

Through a review of 45 studies of FMT in humans and model organisms, we have identified three general criteria that have been applied to assess extent of microbiome engraftment following FMT. The studies we reviewed apply one or more of the following criteria, but we propose that FMT engraftment extent should be quantified with assessments of all three. The first criterion is *Chimeric Asymmetric Community Coalescence*. *Community Coalescence* describes how two microbial communities come together to form a new community[46] In the case of an FMT, the two communities that are merging are the recipient's baseline microbiome and the donated microbiome. *Chimeric* indicates that the communities create an interdependent community network, as opposed to being two independent communities cohabitating.[46] The aspect of the community coalescence that illustrates a successful FMT is that it is *asymmetric,* meaning that the donated microbiome dominates the recipient's microbiome after FMT. To illustrate *Chimeric Asymmetric Community Coalescence*, the microbiome of the recipient after FMT should shift significantly away from their baseline (pre-treatment) microbiome and toward the donated microbiome in terms of richness, composition, and/or structure of the microbiome. The donated microbiome and the recipient's microbiome will almost certainly not be identical.

The second criterion is the presence of *Donated Microbiome Indicator Features*. Specific features present in the donated microbiome, but absent from the recipient pre-transplant, should be observed in the recipient post-transplant. Tracking donated microbiome-associated features in the recipient is one indicator of how effectively the donated microbiome transferred into the recipient following FMT, as opposed to simply introducing a disruption to the recipient's microbiome (which could be confounded with other perturbations such as antibiotic use). Precise tracking of donated features can be difficult with the limited taxonomic resolution of current technologies. Additionally, even if genomically identical organisms were observed in donated microbiome and in the recipient post-transplant, it is effectively impossible to know whether that organism was present in the recipient prior to transplant without sequencing the full genome of every single microbial cell in the recipient's gut prior to transplant. For this reason, we strongly recommend against this being the only criterion used to assess engraftment. Techniques such as strain-profiling with metagenomics data provide higher resolution than, for example, 16S data, but the higher resolution comes at higher cost.

The third and last criterion is *Temporal Stability*. Some methods for microbiome alteration, such as probiotics, often have transient effects on the microbiome[47–49], but if a goal of a treatment is microbiome alteration that persists after treatment ends, it is important that there is short and long-term assessment of the temporal stability in the microbiome shift. A robust FMT should produce a microbiome resistant to reverting back to the baseline microbiome, though long-term similarity to the donated microbiome is not required (and probably not even the goal) as the human gut microbiome is dynamic[34]. Through our literature review we have not identified a consensus for how long that stability should last. FMT studies differ widely in how long and at how many timepoints the microbiome is assessed post-FMT.

It is our perspective that each of these criteria offers complimentary insights, and that all three should be considered in studies that assess the results of FMTs.

## Common Methods for Investigating Engraftment Criteria

Through our literature review, we identified five common methods to evaluate the criteria listed above. In order to assess *Chimeric Asymmetric Community Coalescence*, researchers use one or more of these three common methods: 1) comparing post-transplant alpha diversity to the baseline and/or the donated microbiome, 2) comparing the post-transplant beta diversity distance from the baseline and/or donated microbiome, and 3) source tracking post-transplant features in an attempt to link those features to the donated microbiome. *Donated Microbiome Indicator Features* are assessed using these two common methods: 1) using differential abundance testing techniques to identify features that change in abundance pre- and post-transplant, and 2) tracking previously identified, putatively important features. We did not identify specific approaches for assessing *Temporal Stability*, but rather observed that one or more of the five methods listed above were applied to different post-treatment timepoints. Here we review how these five common methods have been used in FMT studies to investigate our three engraftment extent criteria (Supplemental Data A).

Next, we discuss applications of these criteria in the literature. Most studies use one or more methods to assess engraftment, so citing a study for a specific approach doesn't mean that method was the only approach used. Supplemental Table 1 lists studies using each of the approaches discussed here.

## Criterion 1: Chimeric Asymmetric Community Coalescence

### Alpha Diversity

Alpha diversity estimates within-sample diversity, including community richness and community evenness. Since gut microbiome health is often correlated with higher microbial richness[50–52], richness is often used as a metric of gut microbiome health, although no clear causative relationship between the two has been generally established. A variety of metrics are used to assess alpha diversity. These include richness metrics such as Faith's Phylogenetic Diversity (PD)[53], Observed Features, Simpson Diversity Score[54], and Inverse Simpson Diversity Score[54], Shannon's Index[55], and Pielou's Evenness.[56] To assess FMT engraftment, it is common in the FMT literature to use alpha diversity to compare the recipient to the donated microbiome[57–62], the recipient after FMT intervention to their pre-treatment baseline[30,31,33,34,36,58,61–64], and the recipient to a control.[9,33,34,36,60,65,66] Alpha diversity metrics generally do not capture what features are shared between the donated microbiome and the recipient. For example, two individuals can have the same alpha diversity metric value, but have no features in common.

#### Recipient to Baseline Comparisons

In the reviewed articles, we observed that it was most common to compare the recipient's gut microbiome richness after FMT to their microbiome richness before FMT. If the donated microbiome's richness was higher than the recipient's before FMT (although this information is not always collected), an increase in community richness after FMT is often taken as a sign that the FMT intervention was effective at altering the microbiome.

Post-FMT alpha diversity was significantly increased relative to baseline alpha diversity in many of the studies we reviewed.[30,31,33,34,36,59,61–63,66] Some studies identified no statistically significant change in alpha diversity[67], while others found an apparent increase over time (in some cases statistically significant, but in other cases no statistical test was applied).[37,68–70] Even though most studies show an increase in alpha diversity following transplant, some studies did present a decrease in alpha diversity following transplant.[32,71] For example, Su et al. (2022)[32] tracked community richness of their recipients over the course of their 90 day study and found a significant decrease in evenness as measured by Pielou's Evenness but no significant difference in richness as measured by Observed Features following FMT, which is atypical with a "healthy" donated microbiome. This might be because of the specific diet that was given after the FMT intervention, as the diet-only cohort (who did not receive an FMT) saw an even more drastic decrease in alpha diversity than the FMT group.

Sometimes an individual's baseline sample is already taken after antibiotic use or other treatments, like allo-HCT, which are known to reduce the richness of the gut microbiome. Baseline samples are ideally taken multiple times, including before antibiotics or other treatments [31–34,37,68,69] , but that is not always practical in human subjects research. Amorim et al. (2022)[63] did not measure their baseline prior to antibiotic intervention, and in post-FMT samples found an increase in Shannon's Diversity and Observed Species relative to pre-FMT intervention. In a case like this, it is not possible to determine if an increase in richness is a result of recovery from antibiotic use, the FMT intervention, or a combination of the two. In another example of comparing to a baseline sample after treatment, DeFilipp et al. (2018)[36] showed increased richness by collecting a separate baseline after allo-HCT by using Inverse Simpson Diversity Score to track community richness after FMT intervention relative to two timepoints: before and after allo-HCT treatment.

### Recipient to Donated Microbiome Comparisons

Another common approach for assessing engraftment with alpha diversity is comparing the donated microbiome to the recipient's gut microbiome throughout the FMT study. The purpose of this approach is to track whether the alpha diversity of the recipient is becoming more similar to the donated microbiome over the course of the study.

Some studies showed a significant difference between donated microbiome and the recipients at baseline and then observed recipient community richness become more similar to that of the donated microbiome after FMT.[57–59,72,73] Interestingly, Hazan et al. (2021)[58] used Simpson's and Shannon's diversity indexes and noted that their single patient who did not experience microbiome engraftment had a relatively high community richness prior to FMT intervention, but decreased after FMT intervention. Davar et al. (2021)[12] found no significant differences between the donated microbiome and baseline recipient samples, so they did not use change in alpha diversity as a metric for assessing engraftment.

## Recipient to Control Comparisons

The last common comparison type that we observed involved comparing FMT recipients to a control group that did not receive an FMT, to assess the impact of the intervention.[65,66,72] For example, Kong et al. (2020)[65] used species level change in Shannon's index of their FMT participants compared to their Sham FMT participants (patients who received a placebo). Ma et al. (2023)[72] also compared their findings to a group that was only given PBS. Similarly, Wang et al. (2020)[74] compared FMT recipients at different timepoints to their control group, who received no antibiotics or FMT. Comparisons to controls can elucidate how FMT pre-treatments, like antibiotics, might affect the microbiome[60]. Wang et al. (2020)[74] compared FMT recipients to their spontaneous recovery group, which received antibiotics but no FMT. Another study compared individuals who received allo-HCT and FMT to individuals that received solely allo-HCT treatment.[36] Finally, Kang et al. (2017)[33] and Kang et al. (2019)[34] compared Faith's PD in transplant recipients to an age and gender matched control group.

## Alpha Diversity Visualizations

Alpha diversity metrics are commonly visualized using a boxplot with a scatter plot overlaid (Figure 2 ).[29,32–34,69,70] This helps visualize variation of community richness. Similarly, DeFilipp et al. (2018)[36] used a scatter plot with a median line, and another team used only a scatter plot.[30] Other studies used a trendline, which illustrates how community richness shifts throughout the course of the study (Supplemental Figure 1C).[37,63,65,66,71] Some researchers incorporate a median line illustrating donated microbiome alpha diversity to help illustrate the recipients' alpha diversity compared to the donated microbiomes' (Supplemental Figure 1A).[33,34,37]

Alpha diversity metrics are able to track a shift in microbiome richness and/or evenness, but similar metric values do not indicate that the same features are present. Thus, they can relate microbiome changes to treatment, but they do not provide specific information about the extent to which a microbiome has engrafted. For example, two microbiome samples could have the same *number* of features present, but none of the same features. To tackle more specific questions about whether a recipient's microbiome has become more similar to a donated microbiome in composition, or less similar to their baseline samples in composition, beta diversity metrics are used.

## Beta Diversity

Beta diversity is typically used in microbiome research to quantify the dissimilarity of the taxonomic or phylogenetic microbial composition of pairs of samples. This is often measured as a distance: the larger the distance between a pair of samples, the more dissimilar the samples are to each other. There are a wide variety of distance metrics, including Bray Curtis[76], Jensen–Shannon divergence[77], Unweighted Unifrac distance[78],

Weighted Unifrac distance[78], Jaccard[79], Euclidean distances, Donor Similarity Index[80], Sorensen's Similarity Index[76], and normalized Kimura 2-parameter.[81] Beta diversity methods commonly compare the distance from the FMT recipient's post-FMT gut microbiome to their baseline sample or to the donated microbiome as an assessment of changes induced in the gut microbiome as a result of FMT.

### Distance of Recipient to Donated Microbiome

The approach that we most frequently observed for using beta diversity to assess *Chimeric Asymmetric Community Coalescence* was assessing whether a recipient's distance to the donated microbiome decreased following FMT, by comparing their baseline and post-FMT distances to the donated microbiome. Some studies look at the post-FMT distance to donated microbiome in isolation[60,61,72,73,82], while most track this over multiple timepoints.[12,33,34,36,62,63,65,66,68] Most studies present a significant decrease in distance to donated microbiome sample following FMT.[12,33,34,36,61,63,65,66] For example, Kang et al. (2019)[34] found that Unweighted Unifrac distance to the donated microbiome decreased after administration of the FMT. In another study, Bloom et al. (2022)[68] did not observe a microbial shift towards the donated microbiome's composition with FMT, which might be indicative of engraftment failure, but other methods would help to confirm.

### Distance of Recipient to Baseline

Another common distance calculated is the recipient's distance to their baseline, which is also often tracked over multiple timepoints.[29,39,57,61,64,69,73,83] This comparison helps inform whether the recipient's microbiome after FMT intervention has shifted away from their baseline, another important element of assessing FMT engraftment extent. Gopalakrishnan et al. (2021)[57] used Weighted and Unweighted Unifrac metrics to track distance to baseline over time and noted that recipients who received less frequent doses of FMT (for example, a single dose compared to five doses) remained more similar to their baseline samples at the end of the study relative to recipients who received larger doses, suggesting a dose-dependent response.

### Distance of Recipient to Control

Similarly, some studies compared FMT recipients' distance to a control, such as a Sham FMT group.[63,65,67,72,84,85] This approach was applied to identify how different a placebo group was to an FMT group, which informs whether changes in the microbiome after the transplant are attributable to FMT engraftment, as opposed to other components of the treatment protocol, such as antibiotic use. Wang et al. (2022)[66] used Bray-Curtis dissimilarity index to compare treatment groups to a control that had received antibiotics but no FMT.

### Beta Diversity Visualizations

The most common way to visualize beta diversity in the microbiome literature, and in the studies we reviewed, is through ordination with a method such as Principal Coordinates Analysis (PCoA), and subsequent viewing of the first two or three PCoA axes in a scatter plot (Supplemental Figure 2A).[12,26,27,29,31,32,57,63,66,70,83,86] PCoA plots[87] are convenient approaches for illustrating similarities and differences between groups of samples based on beta diversity distances, frequently highlighting whether groups of samples cluster together (indicating similarity in microbiome composition) by using group-specific sample colors or shapes. In the reviewed studies, sample groups were treatment groups and/or time points. Two studies plotted the distribution of samples along PCoA axis 1 as boxplots (Supplemental Figure 2D)[12,32]. PCoA axis values could also be presented across time in an FMT study using a plot like that presented in Figure 2 (replacing alpha diversity values with the values of specific samples along a single PCoA axis).

While PCoA plots help with visualizing differences in beta diversity on a broad scale, comparing the underlying distances themselves removes a layer of dimensionality reduction. Box plots, typically with jitter plots overlaid, are commonly used to visualize recipients' distances to donated microbiomes (Figure 3A), their baseline samples (Figure 3B, Supplemental Figure 2B), or to control group samples. Multiple box plots can be displayed next to each other to illustrate changes over time or across treatment groups (Figure 3).[33,34,63,66] This allows readers to see the variation of distances to donated microbiomes within groups. As illustrated in Figure 3, distance to the donated microbiome often continues to decrease in the weeks after treatment, suggesting open questions about the microbiome engraftment process and highlighting that summaries such as these don't exclusively present engraftment extent[88]. Another way to visualize distance to donated microbiome is with trendlines tracking recipients' distance from or similarity to a donated microbiome over time, which helps illustrates recipients' individual microbiome shifts (Supplemental Figure 2C).[65,69,83]

## Source Tracking

Tracking how many features and/or the proportion of features that are transferred from the donated microbiome to the recipient can highlight the successful engraftment of the features of interest. One approach for tracking the amount of transferred features is to quantify the proportion of features from varying sources, which is often referred to as source tracking. Often FMT researchers are interested in whether specific features came from a donated microbiome, a recipient, or both.[29,36,37,70] Some FMT studies also track features that come from neither the donated microbiome nor the recipient[36,70], but were possibly recruited by the recipients' diet or other environmental sources after FMT intervention. (Those features could also be sourced from the donated microbiome or recipient, but were present at a level that was not detectable in the relevant source samples.)

Aggarwala et. al. (2021)[89] defined a method for measuring the percentage of donated microbiome features in the recipient's microbiome, Proportional Engraftment of Donor Strains (PEDS). They define this as the total number of donated microbiome strains found in the recipient after FMT invention over the total number of donated microbiome strains.

$$PEDS = \frac{Total\ Number\ of\ Donated\ Microbiome\ Strains\ in\ Recipient}{Total\ Number\ of\ Donated\ Microbiome\ Strains}$$

They also define Proportional Persistence of Recipient Strains (PPRS) as the number of recipient strains that persist in the recipients' microbiome after FMT intervention.

$$PPRS = \frac{Total\ Number\ of\ Recipient\ Strains\ in\ Recipient\ after\ FMT}{Total\ Number\ of\ Recipient\ Strains}$$

In that study, the researchers track the gut microbiome of individuals with recurrent *C. difficile* infections before and after FMT intervention. The mean PEDS in the recipients post-FMT was 75%, and the mean PPRS ranged from 15%-50% for the duration of the 5 year experiment. They reported that 17% PEDS was the threshold for clinical response (ie., no recurrent *C. difficile* infections after FMT intervention).[89] Similarly, Routy et al. (2023) investigated PEDS but defined it as strain engraftment: the number of engrafted strains after FMT divided by the number of strains in the donated microbiome. We would generalize the term "strains" in both of these cases to "features."

SourceTracker[90] is a popular software tool for source tracking of features in microbiome data, where some samples are defined as sources and other samples are defined as sinks. SourceTracker then reports the relative contributions of the different source sample types to each sink. Many studies use methodologies similar to PEDS, PPRS, and SourceTracker, but are custom for the individual study. Of the researchers that defined custom methodologies, the most common method is to label features based on their source and then investigate the source proportions in FMT recipients. The most common labels were donor and recipient, but some contain neither and/or both.[29,36,37,39,70] Gopalakrishnan et al. (2021)[57] slightly modified this by labeling Amplicon Sequence Variant (ASV) sources as mouse, human, or both. They then tracked the percentages of each source in recipient samples throughout their time series. Instead of tracking the proportion over time, Singh et al. (2022)[30] investigated engraftment rate, which they defined as the presence of donated microbiome OTUs found in post-FMT recipient microbiomes that were not observed in the recipient microbiomes at baseline.[30]

Most studies reported an increased percentage of donated microbiome-sourced features in FMT recipients post-FMT relative to baseline. However, Wilson et al. (2021) found that after FMT intervention, there was an increase in features originating from neither the recipient nor the donated microbiome.[70] This indicated that there may be a microbial community developing following the transplant that is not specifically driven by the donated microbiome or recipient's microbiome, but they also note that this could represent a failure to detect low abundance taxa in the donated microbiome or recipient's baseline microbiome.

## Source Tracking Visualizations

The most common source tracking visualizations are trendlines (Supplemental Figure 3B)[37,57,59,89], and barplots (Figure 4A, Supplemental Figure 3A).[29,36,70,89] Trendlines are typically used to understand how the proportions of microbial sources change in the recipient's microbiome over time, while barplots are typically used to understand how specific recipients or groups are changing with respect to microbial source across time (Figure 4A). Scatter plots[30] and boxplots[70] are also used, but they are not as common. Alternatively, a heatmap might be able to capture a subject's PEDS value over time, but only one study used a heatmap to visualize source tracking (Figure 4A).[62]

## Criterion 2: Donated Microbiome Indicator Features

Assessing *Chimeric Asymmetric Community Coalescence* gives a high-level perspective on how the microbiome has changed with FMT. Tracking of specific donated microbiome features is also a popular approach to assessing engraftment that focuses on individual features rather than a summary of the microbiome as a whole. In this case, which we refer to as our *Donated Microbiome Indicator Features* criterion, specific features are considered to serve as indicators of engraftment. We describe two related approaches that differ in how indicator features are identified. Differential abundance (DA) testing, which encompasses a large variety of methods that identify features whose average abundances are significantly different across groups, is central to this approach, but the specific tests that are applied differ.

Linear models are commonly used for testing differential abundance of *Donated Microbiome Indicator Features*.[31,65,68,70,71,83,91] Some teams used linear mixed models[92,93], generalized linear models (MaAsLin2[94])[95], generalized least square models[96], and mixed-effects models.[97] Another team used SourceTracker to predict important species.[59] The time points that investigators compared vary; many teams compared the baseline to time points after the FMT intervention[31,61,64,65,67,68,70,71,73], to the donated microbiome[72,83], to a control group[72,82,84,85,98], or to an alternative FMT method.[99] After DA testing is applied, the identified features are commonly tracked over time as an indicator of engraftment.

El-Salhy et al. (2020)[100] used the GA-map Dysbiosis Test® [101,102], which is based on a predetermined list of features, and then tracked change in this dysbiosis index before and after FMT. Routy et al. (2023), used Aldex2[103] to compare baseline samples to post-FMT-intervention samples.[62] Some researchers used two-tailed t-tests[60,69]. Damman et al. (2015)[69] used paired two-tailed t-tests to compare differences in feature abundance between recipients before FMT intervention and the donated microbiome to see if specific features were missing in the recipient prior to FMT treatment. DA testing on microbiome data is still challenging due to the characteristics of microbiome data, including compositionality and sparsity. Traditional distribution tests, such as t-tests, applied to all features in the microbiome feature table are not a reliable way to identify differentially abundant features. Users of these approaches should assess the state of the field when they are ready to run these analyses to be sure they are using up-to-date methods.

### Tracking previously identified indicator features

A closely related approach is to track microbiome features that were determined to be of relevance to the FMT independently of the current study dataset. The distinction between this approach and those previously described is that the features of interest would be derived from relevant studies on other cohorts (or another source, such as model organism studies). While DA testing could still be applied here, more traditional pairwise comparisons focused on the specific features of interest (rather than all the features observed in a data set) are generally more powerful as they don't require controlling the false discovery rate over very large numbers of comparisons. We refer to this as tracking "previously identified" indicator features, and the types of features used with this approach can vary widely[93].

DeFilipp et al. (2018)[36] investigated *Clostridiales* as an indicator feature after FMT because previous allo-HCT therapy literature showed that *Clostridiales* is a commensal anaerobe that contributes to intestinal homeostasis, and that a decrease of *Clostridiales* had been positively associated with transplant-related mortality.[38,39] They performed a Mann-Whitney-U[104] test on *Clostridiales* relative abundance to test for a significant difference across different timepoints.

Similarly, Kang et al. (2019)[34] tracked changes in relative abundance of three genera: *Bifidobacterium, Prevotella* and *Desulfovibrio. Bifidobacterium* and *Prevotella* are commensal bacteria that were found in previous studies to be depleted in children with autism relative to neurotypical controls.[105,106] *Desulfovibrio* has been reported as both detrimental and commensal in the gut microbiome by different research groups, but has been identified as an important bacterial genus in autism studies.[107] They used a Wilcoxon signed rank test[75] to track changes over time and found that *Desulfovibrio* changes with FMT were significant and *Bifidobacterium* and *Prevotella* were nearly significant.[33] In their two year follow-up study, they compared the $\log_{10}$ of relative abundance of *Bifidobacterium, Prevotella* and *Desulfovibrio* from recipient's baseline sample to time points throughout the experiment using Wilcoxon signed rank test, and they compared recipients to neurotypical controls throughout and after the intervention using Mann-Whitney-U tests.[34]

In another study that tracked previously identified features of interest, the researchers tracked the *Prevotella* to *Bacteroides* ratio, because *Prevotella* and *Bacteroides* were posited to define dominant "enterotypes" that are helpful for clinical diagnoses.[108] (The enterotype hypothesis is disputed due to significant variation and instability within defined enterotypes, and no distinct boundaries between enterotypes.[109–111]) The team found that the recipients did have an increased *Prevotella* to *Bacteroides* ratio after FMT intervention.[70]

Lastly, van Lier et al. (2020)[37] tracked predicted butyrate-producing bacteria as well as *Blautia* and *Clostridiales* across FMT study timepoints. They found that clinical responders to the FMT seemed to have an increase in *Clostridiales* and other predicted butyrate-producing bacteria relative to the clinical non-responders.[37]

## Features Abundance Tracking Visualizations

DA can be complex to visualize because most differential abundance methods that are relevant to microbiome data transform the data for testing. This means that teams have to visualize using either transformed data (Figure 5A), which can be hard to interpret, or relative abundance (Figure 5B), which is easier to interpret but may not clearly reflect important aspects of the test. Boxplots were one of the most common visualizations in the discussed studies (Supplemental Figure 4C).[33,34,68,70,83] Similarly, Parker et al. (2022)[31] used a diverging barplot with mean differences in centered log ratio and standard error (Supplemental Figure 4A).[31] Another common visualization is a heat map of relative abundance (Supplemental Figure 4B)[70], or normalized distance.[65] Trendlines are also used to track relative abundance of the differentially abundant species over time (Figure 5B)[37,83], while other teams use scatter plots.[26,36]

## Criterion 3: Temporal Stability

Stability of engraftment is an important aspect of FMT intervention. Clinicians often have a checkup with the patient 3-7 days after FMT intervention, and another checkup at 4-8 weeks is recommended[20], and it can be useful to use these opportunities to assess engraftment extent by testing whether the criteria described above have sustained over time. In probiotic studies, there are often transient features, which might benefit the recipient temporarily due to their metabolic activity, but offer limited long-term changes to the microbiome.[15,47–49] However, the goal of an FMT is generally the transition away from an unhealthy microbiome state toward a new-to-the-recipient healthier microbiome state. The changes introduced by the FMT should therefore have some degree of stability. Most FMT studies are longitudinal and track the microbiome over time to investigate the FMT's effects after the initial intervention. However, there is no agreement on how long the microbiome needs to be maintained after FMT intervention to qualify as engraftment. The median study length was 67.5 days (among papers that directly stated the study length, n=40) between the first and last time point sample. The minimum length of these studies was 18 days and the maximum length was 1,825 days (Supplemental Data B). Minimal justification for study length was given. Standardization of the FMT follow-up timeline would help to compare FMT engraftment extent across studies. Kang et al. (2019) saw that after two years, the microbiome of the recipients did not look like either the donated microbiome or participants' baseline microbiome, as assessed by unweighted UniFrac and other beta diversity metrics, but rather looked like a new microbiome state relative to those previously collected in the study. The authors still considered the engraftment to have been successful, because the recipients retained an increased community richness at the two year follow-up, and their richness was no longer significantly different from controls, as it was at baseline. This suggests that the recipients maintained the higher community richness that was induced by the FMT, but that their microbiome was again "personalized," which might be due to environmental factors.[34] While not always possible or practical, long-term follow-up in FMT studies can highlight outcomes such as this, which can support protocols for safer FMTs and better understanding of the impact of FMTs on the human gut microbiome.

## Longitudinal Statistics

The majority of FMT studies collect microbiome samples longitudinally. Repeated measures over time introduce data analysis challenges, including dependent data, irregularly timed data, missing timepoints, and dropped subjects.[112]

Dependence of data is one of the biggest issues when it comes to analyzing longitudinal data. Many statistical tests assume that the data points are independent, and repeat measures from the same subject violate this assumption. This limits the statistical tests that can be applied, and using tests that assume independence is a common error. Further, comparing timepoints to each other without pairing samples from the same subject can reduce statistical power for assessing the impact of an FMT or other microbiome intervention. For example, Figure 6 shows that the groups "Baseline" and "Post-FMT" might not be statistically different based on how the groups cluster together (Figure 6A). However, if change between the adjacent timepoints is calculated on a per-subject basis, sometimes referred to as "first distance," it highlights that each time point shifted towards the donated microbiome (Figure 6B). Thus, applying the appropriate statistical test can highlight important outcomes that otherwise would be obscured.[113] Another technique for analyzing longitudinal data that was common in the reviewed studies was with linear modeling techniques that support tracking change over time and correlations with other variables.[114] Appropriately constructed models can control the effects of correlated data or irregular time points.[11,112]

# Conclusions

In this manuscript, we summarize the methods and approaches applied in recent literature for quantifying FMT engraftment, and categorize them into three criteria. The first criterion, which we refer to here as *Chimeric Asymmetric Community Coalescence,* provides a general overview of how the microbiome changes after FMT intervention. The three approaches we identified for assessing *Chimeric Asymmetric Community Coalescence* use alpha diversity, beta diversity, or microbial source tracking. Alpha diversity can describe whether the microbiome richness or evenness changes after FMT, but does not show the changes occuring in microbiome composition. Beta diversity assesses whether the microbiome composition of the recipient shifted towards the donated microbiome and/or away from the recipient's baseline microbiome, based on a community-wide overview. Source tracking methods track the proportion of donated microbiome features that successfully engrafted into the recipient. None of these three metrics confirm that specific donated microbiome features have successfully engrafted. The second criterion, *Donated Microbiome Indicator Features,* captures how specific feature abundance changes with FMT intervention. These approaches do not capture the changes in the microbiome composition as a whole, and can be impacted by low abundance features falling below levels of detection. This criterion can be explored by analyzing the differential abundance of features before and after

FMT, ideally using paired samples. Features of interest can be either derived from the current study samples or pre-determined independently of the current study dataset. The third criterion, *Temporal Stability,* captures whether the changes in the recipient's microbiome following FMT are resistant to reverting back to the baseline microbiome. This criterion does not have any additional metrics associated with it based on the literature that we reviewed, but rather *Temporal Stability* tends to be investigated using any of the methods listed above, as long as they are assessing the microbiome longitudinally at more than two post-FMT timepoints.

Taken together, these three criteria of FMT engraftment assess how the microbiome shifted at a high level, what features were important in that shift, and the stability of the shift. However, these criteria, on their own, do not completely answer the question of whether the microbiome engrafted. Rather, each method provides a different view of microbiome engraftment. We recommend using approaches that address all three of these criteria to obtain the clearest insight into FMT engraftment.

In the following sections, we discuss additional considerations in assessing microbiome engraftment with FMT.

## Microbiome data type

Most studies reviewed here applied these methods to 16S amplicon data, but many of the same methods could be applied to metagenomics data (e.g., to assess the functional potential of the transplanted microbiome and/or enable higher taxonomic resolution variants of the methods highlighted here[66,70,115]), metatranscriptomics and/or metaproteomics data (to assess whether transplanted microbiomes perform the same functions in the recipients as they do in the hosts), or even metabolomics data (e.g., to assess change in postbiotics). Increased taxonomic resolution through microbiome metagenomics enables tracking specific donated microbiome features with higher specificity, for example at the species or strain level. Additionally, if an FMT benefits the recipient based on the beneficial functional traits that are transferred, not the specific microbes[115], then metagenomics data would assess functional trait transfer better. However, metagenomics can be problematic in low microbial biomass samples due to human DNA contamination, and the higher sequencing costs sometimes compromise study design by requiring infrequent temporal sampling.

Recently, multiple meta-analyses have discussed strain engraftment with metagenomic data. Podlesny et al. (2022)[116] found that donated microbiome strain engraftment is highly variable based on pre-treatments, alpha diversity of recipients/donors, abundances of species, and functional traits of species. Schmidt et al. (2021)[117] found that donated microbiome strain engraftment and recipient strain resilience were not strongly correlated with clinical outcomes. However, Ianiro et al. (2022)[118] found that donated microbiome strain engraftment was highly correlated with clinical response, but also noted that administration frequency and antibiotics increased engraftment extent.

## Phylogenetic and Quantitative Diversity Metrics in FMT Engraftment Assessment

As far as we can tell, phylogenetic diversity metrics have not been applied in the context of FMT to compute a metric similar to PEDS, even though phylogenetic methods generally are important in microbiome data science. UniFrac Gain[119], or a related asymmetric dissimilarity measure, could be applied to achieve this purpose, reporting the percentage of phylogenetic branch length represented in a donor sample that is newly observed in a recipient following FMT. However when quantifying change with FMT by tracking features, we are interested in knowing whether specific features transferred, as opposed to closely related features. For this reason, tracking identical features at the maximum resolution allowed by the technology being applied, rather than weighting features by their phylogenetic relatedness, is likely more informative for quantifying microbiome engraftment.

Similarly, qualitative rather than quantitative diversity metrics (such as Jaccard distance versus Bray-Curtis dissimilarly) may provide more direct insight into engraftment extent. Again, when the goal is to transfer specific features, and not necessarily a change in abundance of features, then methods that compare the presence or absence of features are more closely aligned with the goal of determining if specific features were transferred.

## Comparing Across Studies

Comparing engraftment results across FMT studies is difficult because the different metrics used are generally not directly comparable across studies. For example, alpha diversity metrics such as Faith's PD and Observed Features are difficult to directly compare because they have different meanings: Faith's PD includes phylogenetic distances and Observed Features does not. Even comparing the same metric across studies can be difficult because results are impacted by differences in primer choices, depth of sequencing, and sequencing quality control approaches (and phylogenetic metrics are additionally impacted by the approach used for creating the phylogenetic tree). As a result, comparing microbiome engraftment extent across studies tends to be qualitative, focusing on high-level outcomes like "richness increased following FMT" or "distance to donated microbiome decreased following FMT". Creating standardized methods to assess engraftment is a first step toward addressing this problem and enabling for more direct comparison across studies that quantify engraftment.

## Guidelines for Sample Collection

Sample collection strategies and timing varied widely in the studies reviewed here, including which types of samples that were collected, frequency of sampling, and duration of sampling. Here we provide guidelines for samping to best enable assessment of engraftment. Of course, when working with human subjects, particularly those who are already burdened with illness, achieving these ideals may not be always feasible.

At a bare minimum, to assess change with FMT, samples must be collected before and after the FMT. We also consider it critical to sample the donated microbiome. With these three samples per subject, it is possible to track Asymmetric Chimeric Community Coalescence, by testing whether the recipient's microbiome is more similar to the donated microbiome following treatment, and the transfer of Donated Microbiome Indicator Features, by checking for the presence of features that are observed in both the donated microbiome and the recipient's microbiome post-FMT, but which are not present in the recipient's microbiome pre-FMT. If a recipient receives multiple doses of FMT from different sources (as in Kang et al. (2017))[33], each of the sources should be sequenced.

Ideally, multiple samples will be collected from the recipient prior to and after the FMT beginning as soon as possible after FMT intervention (e.g., first bowel movement following FMT) so that low engraftment extent at the beginning of the procedure could be identified and potentially rectified. Multiple pre-FMT timepoints provides a more representative view of the recipient's microbiome as normal temporal variation can be averaged, and because there are usually pre-FMT treatments, such as antibiotics, and it can be informative to track change across those treatments to conclusively link a change in the microbiome to an FMT treatment (rather than pre-treatment antibiotics, for example). Multiple timepoints following FMT is necessary for understanding temporal stability of the transplant.

It is not currently clear what length of time samples should be collected for following FMT. Additional timepoints, where practical, won't ever hurt the data analysis. Collecting and reporting on post-FMT microbiomes will aid in our understanding of what range of stability should be expected, what range of stability should be considered abnormal, and whether stability (or lack thereof) is associated with longer term clinical outcomes.

## So, did the microbiome engraft?

Assessing whether a donated microbiome engrafted following FMT, and relating that information to clinical outcomes, remains challenging. In part, this is because the suitable answer to the question of engraftment is not "yes" or "no": we argue that a quantitative metric of engraftment extent is more appropriate than a binary measure of engraftment success. Additionally, microbiome functional activity transfer is almost certainly more relevant to clinical outcomes than microbiome composition or functional potential transfer, but activity is considerably more challenging to measure.

Currently, we define high engraftment extent as meeting the three criteria we defined in this manuscript, low engraftment extent as meeting one or two of these criteria, and no engraftment as not meeting any of these criteria. We do not attempt to define an exact threshold for engraftment extent. Further, a lack of standard methods, the challenges with comparing measures of engraftment extent across studies, and few direct assessments linking success of FMT with clinical outcomes, make it unclear whether crossing (currently undefined) thresholds of engraftment extent is linked with positive clinical outcomes.

To move FMT research forward, software tools should be developed that make it straightforward for different research teams to assess all of the criteria outlined here using the same approaches, and these assessments should ultimately be translated into clinician-targeted reports. Routinely relating engraftment extent based on these criteria to clinical outcomes in FMT research and praxis will allow us to determine if some engraftment extent measures are more useful than others, and to develop an understanding of what extent of engraftment should be targeted to optimize clinical outcomes.

# Figures

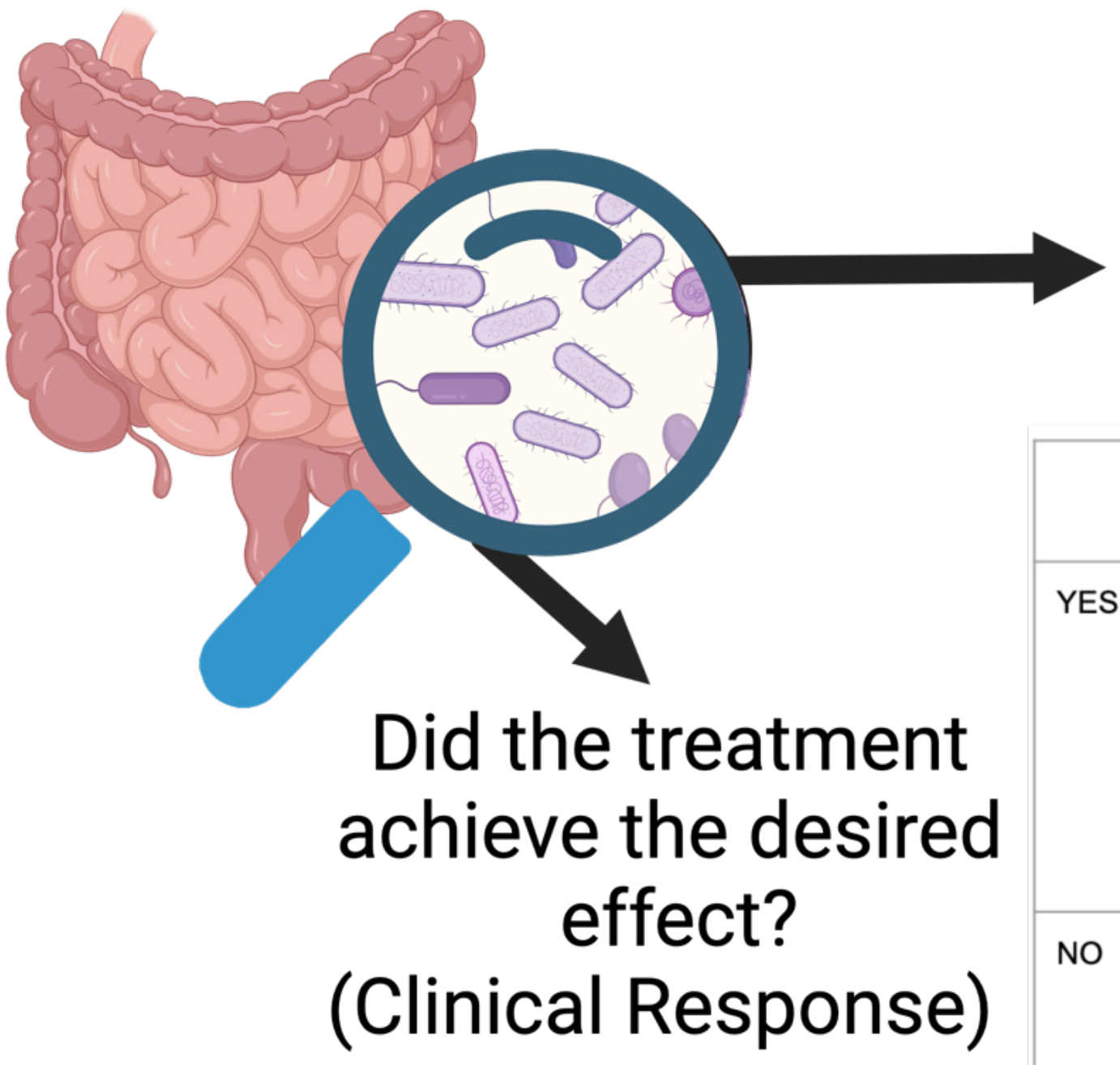

Did the microbiome engraft?
(Engraftment Extent)

| | YES | NO |
|---|---|---|
| YES | FMT may be associated with positive treatment outcome | Positive treatment outcome may be associated with confounding factors |
| NO | FMT may not be associated with positive treatment outcome | No clear insight gained |

**Figure 1: Assessing the success of FMT.** When investigating outcomes of an FMT, we argue that two questions should be asked: did the microbiome engraft, and did the treatment achieve the desired outcome? Created with BioRender.com.

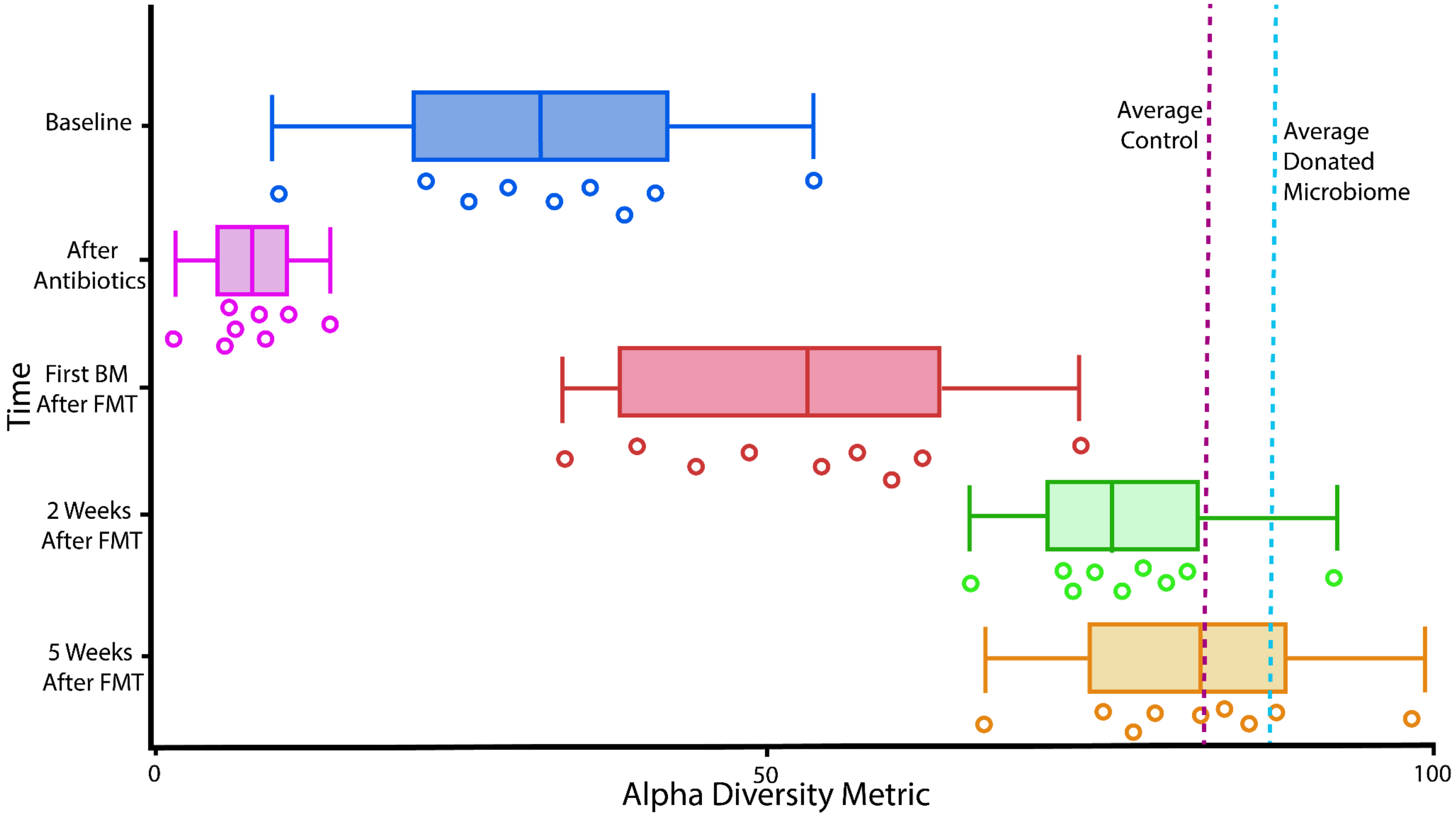

**Figure 2: Schematic of Chimeric Asymmetric Community Coalescence assessment with an alpha diversity metric.** This visualization summarizes alpha diversity distributions at each time point with box plots, provides additional detail on the distributions in the corresponding jitter plots, and the *Average Control* and *Average Donated Microbiome* reference lines aid in contextualizing the metric. With appropriate statistics, such as the Wilcoxon signed rank test[75], this plot could be used to suggest engraftment in recipients as assessed by Chimeric Asymmetric Community Coalescence with an alpha diversity metric, but we note again the caveat that similar measures of alpha diversity could be achieved with no shared features, so we do not consider this to be strong evidence of engraftment. *First BM After FMT Treatment* indicates the recipient's first bowel movement (BM) following the fecal microbiota transplant.

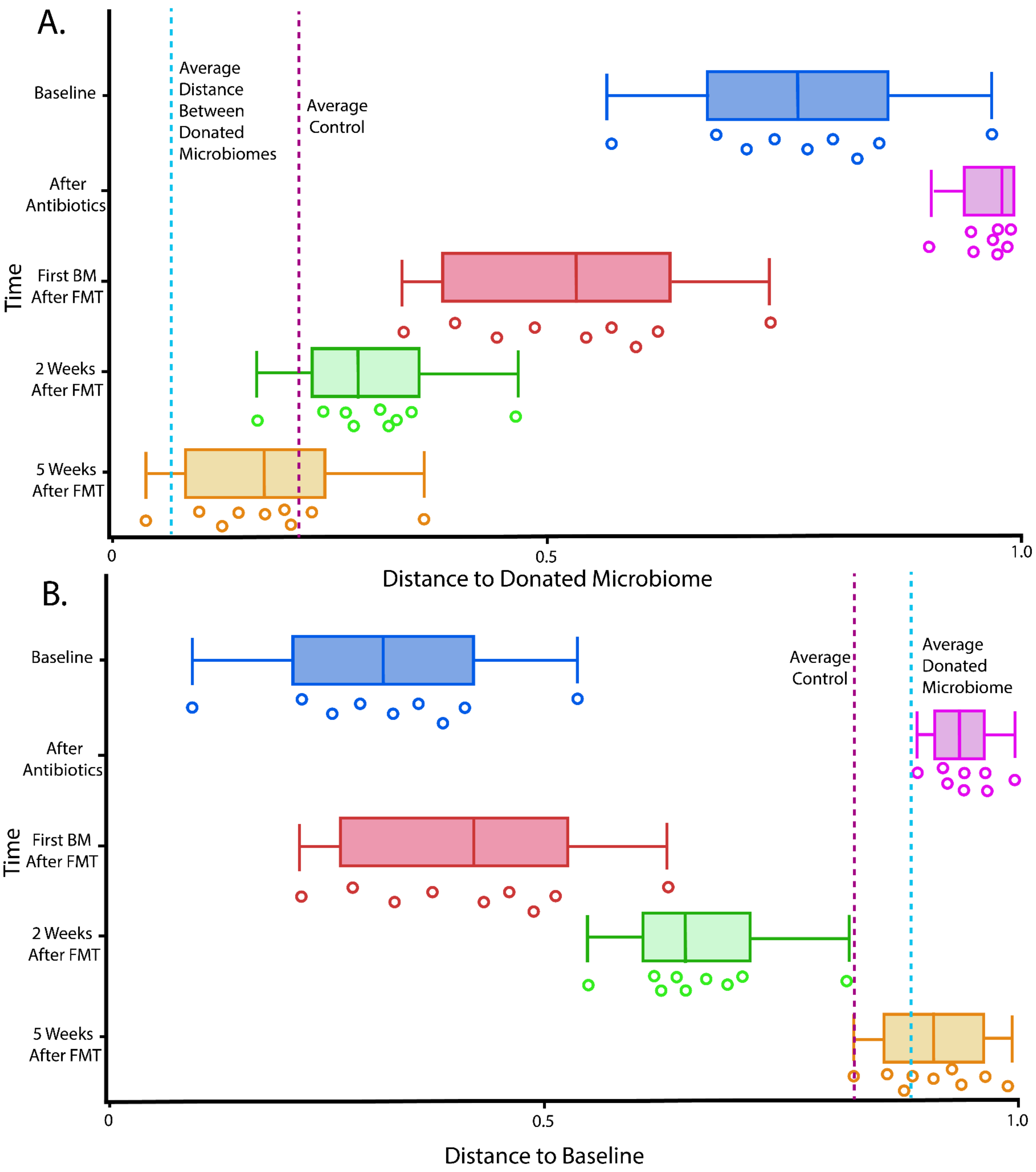

**Figure 3: Schematic of *Chimeric Asymmetric Community Coalescence* assessment with beta diversity distances between samples**. This visualization illustrates distances to (A) donated microbiomes and (B) individuals' baseline samples. Relevant average control and donor distance reference lines aid in contextualizing distances. With appropriate statistics, such as Wilcoxon signed rank test, panel A could be used to suggest engraftment based on a decreasing distance to the donated microbiome samples with treatment and panel B could be used to suggest engraftment based on increasing distance from baseline samples with treatment. *First BM After FMT Treatment* indicates the recipient's first bowel movement (BM) following the fecal microbiota transplant.

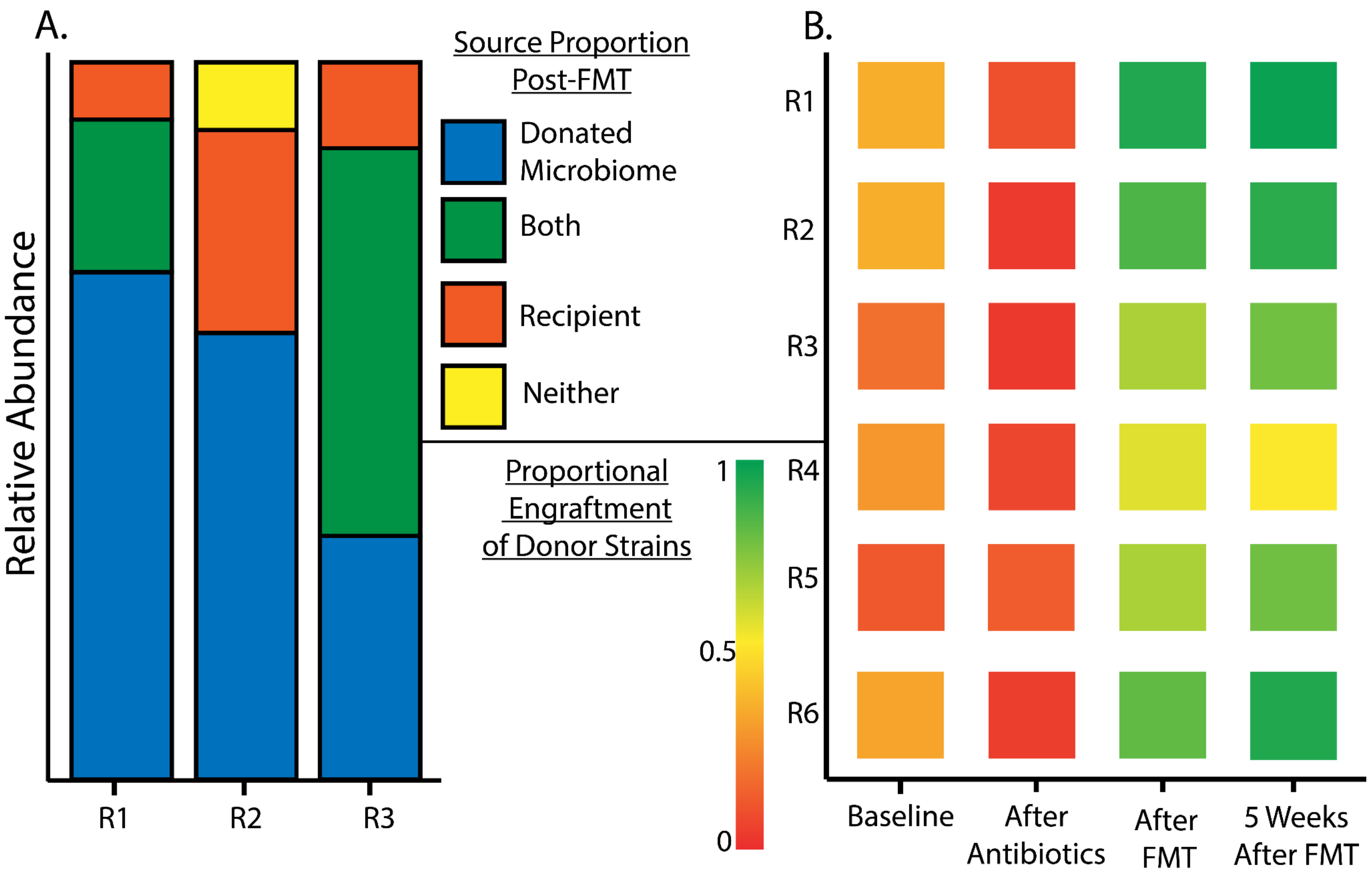

**Figure 4: Schematic of *Chimeric Asymmetric Community Coalescence* assessment with microbiome feature source tracking.** (A) Recipient microbiomes are described at a single post-FMT timepoint in terms of the source of each feature in the microbiome: the donated microbiome, the recipient's baseline microbiome (recipient), both, or neither. This could be generalized to provide this information across multiple timepoints as well. (B) PEDS is tracked across time and across subjects, illustrating larger values post-FMT relative to baseline. R1 - R6 refer to 6 theoretical recipients of an FMT.

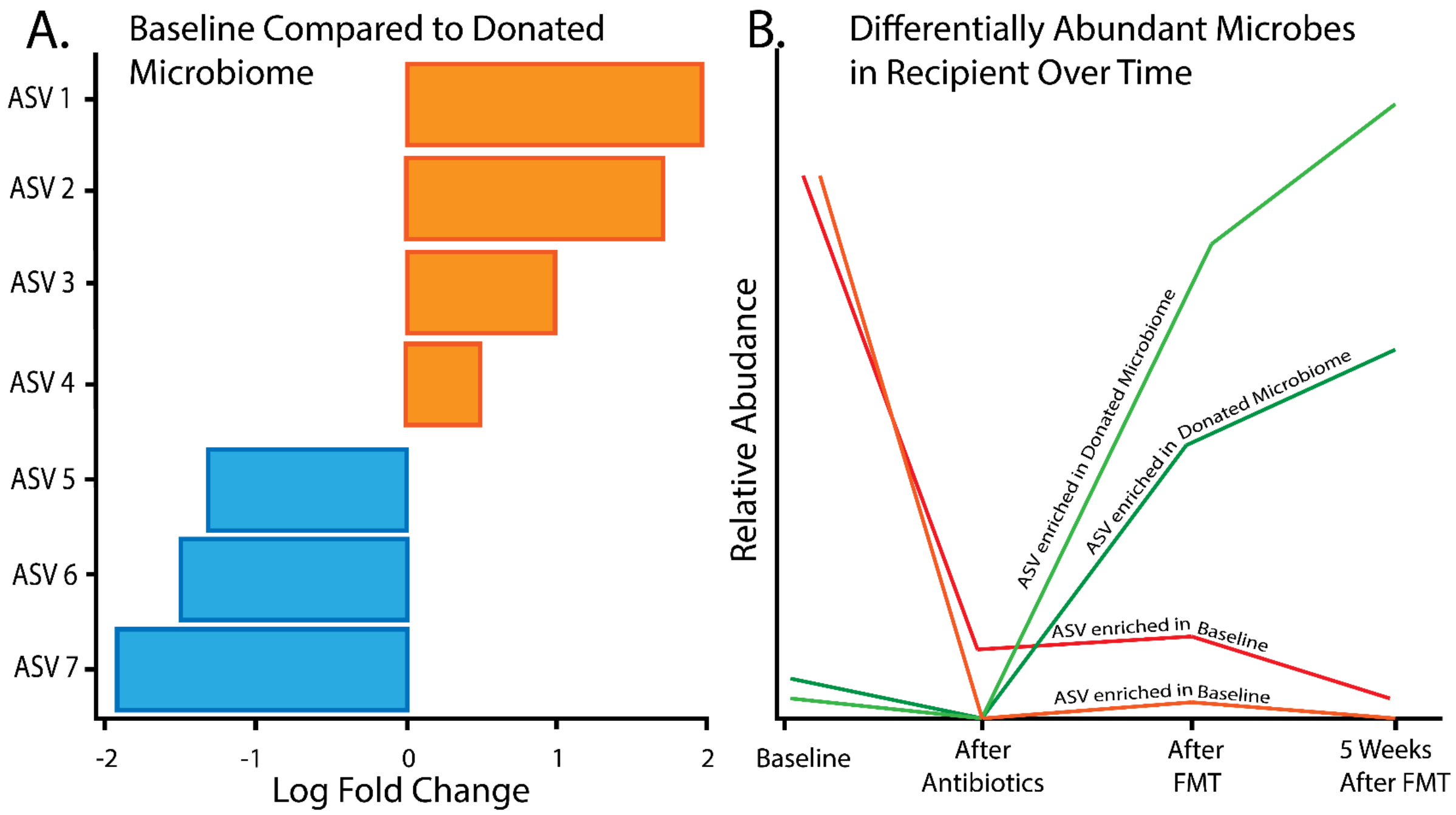

**Figure 5: Feature Abundance Tracking Visualization.** This visualization contains **(A)** diverging bar plot with log fold change to illustrate differentially abundant microbes between the donated microbiome and baseline. Orange bars indicate ASVs enriched in the baseline compared to donated microbiomes, while blue bars indicate ASVs that are depleted in the baseline compared to donated microbiomes. Subsection **(B)** if the visualization tracks how the relative abundance changes over time.

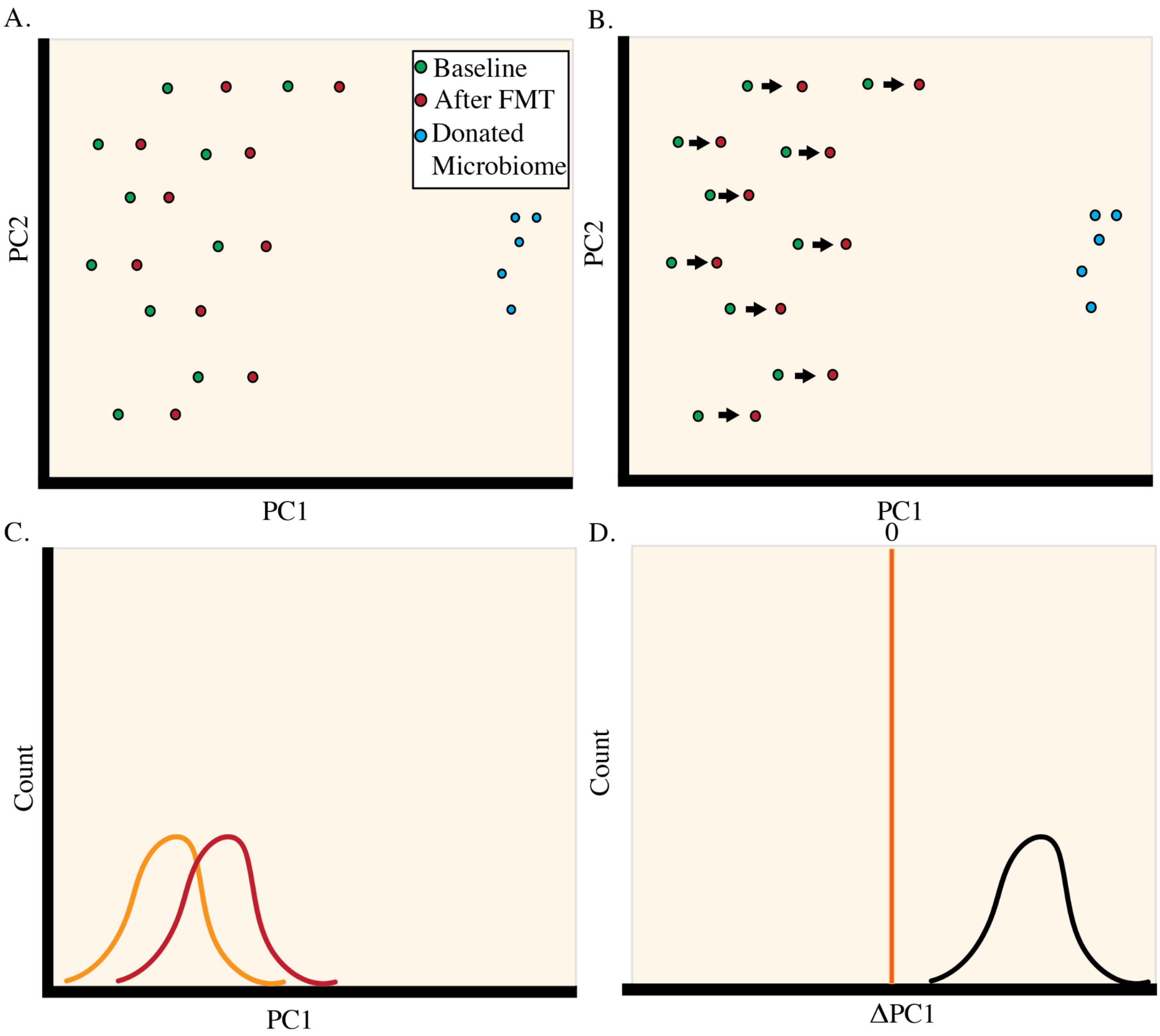

**Figure 6: First Distances Capture Rate of Change.** A) Baseline and After FMT groups do not look like distinct microbiomes. B) Looking at first distances captures that each subject shifted to towards the donated microbiome on the PC Axis 1. C) Distributions of baseline and after FMT samples, as described by PC, overlap, indicating probable no significant difference. D) The distribution of change between baseline and after FMT samples as described by PC1 does not overlap with the null hypothesis, indicating a probable significant change. *This figure was adapted from ASM Microbe Volume 9, Number 10, 2014 with permission from ASM.*

# Declarations

## Ethics approval and consent to participate

Not applicable (not human subjects or animal research).

## Consent for publication

Not applicable (not human subjects research).

## Funding

This work was funded in part by NIH National Cancer Institute Informatics Technology for Cancer Research Award 1U24CA248454-01 to JGC.

## Availability of data and materials

Not applicable.

## Competing interests

The authors declare that they have no competing interests.

## Authors' contributions

C.H. and J.G.C wrote the main manuscript. B.M.B. advised on terminology used for defining FMT engraftment extent criteria. T.F.B., V.C., R.K-B., L.L., C.L., F.M., K.N., and J.B.X. advised on relevance of engraftment extent criteria in FMT research praxis, and assisted with defining engraftment extent criteria. M.J. provided editing assistance throughout the manuscript development and review process, and assisted with defining engraftment extent criteria. All authors reviewed drafts of the manuscript and provided feedback during its development.

## Acknowledgements

Not applicable.

# List of Works Cited

1. Zheng, D., Liwinski, T. & Elinav, E. Interaction between microbiota and immunity in health and disease. *Cell Res.* **30**, 492–506 (2020).
2. Li, Y. *et al.* Effects of Gut Microbiota on Host Adaptive Immunity Under Immune Homeostasis and Tumor Pathology State. *Front. Immunol.* **13**, (2022).
3. Ferreira, C. M. *et al.* The Central Role of the Gut Microbiota in Chronic Inflammatory Diseases. *Journal of Immunology Research* **2014**, 689492 (2014).
4. Glassner, K. L., Abraham, B. P. & Quigley, E. M. M. The microbiome and inflammatory bowel disease. *J. Allergy Clin. Immunol.* **145**, 16–27 (2020).
5. Wlodarska, M., Kostic, A. D. & Xavier, R. J. An Integrative View of Microbiome-Host Interactions in Inflammatory Bowel Diseases. *Cell Host Microbe* **17**, 577–591 (2015).
6. Wu, H.-J. & Wu, E. The role of gut microbiota in immune homeostasis and autoimmunity. *Gut Microbes* **3**, 4–14 (2012).
7. Xu, H. *et al.* The Dynamic Interplay between the Gut Microbiota and Autoimmune Diseases. *Journal of Immunology Research* **2019**, 7546047 (2019).
8. Miyauchi, E., Shimokawa, C., Steimle, A., Desai, M. S. & Ohno, H. The impact of the gut microbiome on extra-intestinal autoimmune diseases. *Nat. Rev. Immunol.* **23**, 9–23 (2023).
9. Yu, H. *et al.* Fecal microbiota transplantation inhibits colorectal cancer progression: Reversing intestinal microbial dysbiosis to enhance anti-cancer immune responses. *Front. Microbiol.* **14**, 1126808 (2023).
10. Nandi, D., Parida, S. & Sharma, D. The gut microbiota in breast cancer development and treatment: The good, the bad, and the useful! *Gut Microbes* **15**, 2221452 (2023).
11. Taur, Y. *et al.* Reconstitution of the gut microbiota of antibiotic-treated patients by autologous fecal microbiota transplant. *Sci. Transl. Med.* **10**, eaap9489 (2018).
12. Davar, D. *et al.* Fecal microbiota transplant overcomes resistance to anti–PD-1 therapy in melanoma patients. *Science* **371**, 595–602 (2021).

13. Borsom, E. M. *et al. Predicting Neurodegenerative Disease Using Pre-Pathology Gut Microbiota Composition: A Longitudinal Study in Mice Modeling Alzheimer''s Disease Pathologies*. https://www.researchsquare.com/article/rs-1538737/v1 (2022) doi:10.21203/rs.3.rs-1538737/v1.

14. Cryan, J. F., O'Riordan, K. J., Sandhu, K., Peterson, V. & Dinan, T. G. The gut microbiome in neurological disorders. *Lancet Neurol.* **19**, 179–194 (2020).

15. Suganya, K. & Koo, B.-S. Gut–Brain Axis: Role of Gut Microbiota on Neurological Disorders and How Probiotics/Prebiotics Beneficially Modulate Microbial and Immune Pathways to Improve Brain Functions. *Int. J. Mol. Sci.* **21**, 7551 (2020).

16. Dinan, T. G. & Cryan, J. F. The Microbiome-Gut-Brain Axis in Health and Disease. *Gastroenterol. Clin. North Am.* **46**, 77–89 (2017).

17. Cryan, J. F. & Dinan, T. G. Mind-altering microorganisms: the impact of the gut microbiota on brain and behaviour. *Nat. Rev. Neurosci.* **13**, 701–712 (2012).

18. Gebrayel, P. *et al.* Microbiota medicine: towards clinical revolution. *J. Transl. Med.* **20**, 111 (2022).

19. Zhang, Y., Zhou, L., Xia, J., Dong, C. & Luo, X. Human Microbiome and Its Medical Applications. *Frontiers in Molecular Biosciences* **8**, (2022).

20. Kim, K. O. & Gluck, M. Fecal Microbiota Transplantation: An Update on Clinical Practice. *Clin. Endosc.* **52**, 137–143 (2019).

21. Hui, W., Li, T., Liu, W., Zhou, C. & Gao, F. Fecal microbiota transplantation for treatment of recurrent C. difficile infection: An updated randomized controlled trial meta-analysis. *PLoS One* **14**, e0210016 (2019).

22. Weiss, G. A. & Hennet, T. Mechanisms and consequences of intestinal dysbiosis. *Cell. Mol. Life Sci.* **74**, 2959–2977 (2017).

23. Olesen, S. W. & Alm, E. J. Dysbiosis is not an answer. *Nat Microbiol* **1**, 16228 (2016).

24. Commissioner, O. of T. FDA Approves First Fecal Microbiota Product. https://www.fda.gov/news-events/press-announcements/fda-approves-first-fecal-microbiota-product (2022).

25. Office of the Commissioner. FDA Approves First Orally Administered Fecal Microbiota Product for the Prevention of Recurrence of Clostridioides difficile Infection. *U.S. Food and Drug Administration*

https://www.fda.gov/news-events/press-announcements/fda-approves-first-orally-administered-fecal-microbiota-product-prevention-recurrence-clostridioides (2023).

26. Baruch, E. N. *et al.* Fecal microbiota transplant promotes response in immunotherapy-refractory melanoma patients. *Science* **371**, 602–609 (2021).

27. El-Salhy, M. *et al.* Long-term effects of fecal microbiota transplantation (FMT) in patients with irritable bowel syndrome. *Neurogastroenterology & Motility* **34**, e14200 (2022).

28. Wang, Y. *et al.* Fecal microbiota transplantation for refractory immune checkpoint inhibitor-associated colitis. *Nat. Med.* **24**, 1804–1808 (2018).

29. Riquelme, E. *et al.* Tumor Microbiome Diversity and Composition Influence Pancreatic Cancer Outcomes. *Cell* **178**, 795–806.e12 (2019).

30. Singh, P. *et al.* Effect of antibiotic pretreatment on bacterial engraftment after Fecal Microbiota Transplant (FMT) in IBS-D. *Gut Microbes* **14**, 2020067.

31. Parker, A. Fecal microbiota transfer between young and aged mice reverses hallmarks of the aging gut, eye, and brain. 25 (2022).

32. Su, L. *et al.* Health improvements of type 2 diabetic patients through diet and diet plus fecal microbiota transplantation. *Sci. Rep.* **12**, 1152 (2022).

33. Kang, D.-W. *et al.* Microbiota Transfer Therapy alters gut ecosystem and improves gastrointestinal and autism symptoms: an open-label study. *Microbiome* **5**, 10 (2017).

34. Kang, D.-W. *et al.* Long-term benefit of Microbiota Transfer Therapy on autism symptoms and gut microbiota. *Sci. Rep.* **9**, 5821 (2019).

35. Park, S.-H. *et al.* Cognitive function improvement after fecal microbiota transplantation in Alzheimer's dementia patient: a case report. *Curr. Med. Res. Opin.* **37**, 1739–1744 (2021).

36. DeFilipp, Z. *et al.* Third-party fecal microbiota transplantation following allo-HCT reconstitutes microbiome diversity. *Blood Advances* **2**, 745–753 (2018).

37. van Lier, Y. F. *et al.* Donor fecal microbiota transplantation ameliorates intestinal graft-versus-host disease in allogeneic hematopoietic cell transplant recipients. *Sci. Transl. Med.* **12**, eaaz8926 (2020).

38. Jenq, R. R. *et al.* Regulation of intestinal inflammation by microbiota following allogeneic bone marrow transplantation. *J. Exp. Med.* **209**, 903–911 (2012).

39. Taur, Y. *et al.* The effects of intestinal tract bacterial diversity on mortality following allogeneic hematopoietic stem cell transplantation. *Blood* **124**, 1174–1182 (2014).

40. Gopalakrishnan, V., Helmink, B. A., Spencer, C. N., Reuben, A. & Wargo, J. A. The Influence of the Gut Microbiome on Cancer, Immunity, and Cancer Immunotherapy. *Cancer Cell* **33**, 570–580 (2018).

41. Fong, W., Li, Q. & Yu, J. Gut microbiota modulation: a novel strategy for prevention and treatment of colorectal cancer. *Oncogene* **39**, 4925–4943 (2020).

42. Xu, F. *et al.* The efficacy of prevention for colon cancer based on the microbiota therapy and the antitumor mechanisms with intervention of dietary Lactobacillus. *Microbiol Spectr* **11**, e0018923 (2023).

43. Winn, A. N., Ekwueme, D. U., Guy, G. P., Jr & Neumann, P. J. Cost-Utility Analysis of Cancer Prevention, Treatment, and Control: A Systematic Review. *Am. J. Prev. Med.* **50**, 241–248 (2016).

44. Callahan, B. J., McMurdie, P. J. & Holmes, S. P. Exact sequence variants should replace operational taxonomic units in marker-gene data analysis. *ISME J.* **11**, 2639–2643 (2017).

45. Khanna, S. *et al.* Efficacy and Safety of RBX2660 in PUNCH CD3, a Phase III, Randomized, Double-Blind, Placebo-Controlled Trial with a Bayesian Primary Analysis for the Prevention of Recurrent Clostridioides difficile Infection. *Drugs* **82**, 1527–1538 (2022).

46. Custer, G. F., Bresciani, L. & Dini-Andreote, F. Ecological and Evolutionary Implications of Microbial Dispersal. *Front. Microbiol.* **13**, 855859 (2022).

47. Hemarajata, P. & Versalovic, J. Effects of probiotics on gut microbiota: mechanisms of intestinal immunomodulation and neuromodulation. *Therap. Adv. Gastroenterol.* **6**, 39–51 (2013).

48. Wieërs, G. *et al.* How Probiotics Affect the Microbiota. *Front. Cell. Infect. Microbiol.* **9**, 454 (2020).

49. Grond, K. *et al.* Longitudinal microbiome profiling reveals impermanence of probiotic bacteria in domestic pigeons. *PLoS One* **14**, e0217804 (2019).

50. Lozupone, C. A., Stombaugh, J. I., Gordon, J. I., Jansson, J. K. & Knight, R. Diversity, stability and resilience of the human gut microbiota. *Nature* **489**, 220–230 (2012).

51. Mosca, A., Leclerc, M. & Hugot, J. P. Gut Microbiota Diversity and Human Diseases: Should We Reintroduce Key Predators in Our Ecosystem? *Front. Microbiol.* **7**, (2016).

52. Scepanovic, P. *et al.* A comprehensive assessment of demographic, environmental, and host genetic associations with gut microbiome diversity in healthy individuals. *Microbiome* **7**, 130 (2019).

53. Faith, D. P. Conservation evaluation and phylogenetic diversity. *Biol. Conserv.* **61**, 1–10 (1992).

54. Simpson, E. H. Measurement of Diversity. *Nature* **163**, 688–688 (1949).

55. Shannon, C. E. A mathematical theory of communication. *The Bell System Technical Journal* **27**, 379–423 (1948).

56. Pielou, E. C. The measurement of diversity in different types of biological collections. *J. Theor. Biol.* **13**, 131–144 (1966).

57. Gopalakrishnan, V. *et al.* Engraftment of Bacteria after Fecal Microbiota Transplantation Is Dependent on Both Frequency of Dosing and Duration of Preparative Antibiotic Regimen. *Microorganisms* **9**, 1399 (2021).

58. Hazan, S., Dave, S., Papoutsis, A. J., Barrows, B. D. & Borody, T. J. Successful Bacterial Engraftment Identified by Next-Generation Sequencing Predicts Success of Fecal Microbiota Transplant for Clostridioides difficile. *Gastroenterol. Res. Pract.* **14**, 304–309 (2021).

59. Staley, C. *et al.* Durable Long-Term Bacterial Engraftment following Encapsulated Fecal Microbiota Transplantation To Treat Clostridium difficile Infection. *MBio* **10**, e01586–19 (2019).

60. Zeng, X. *et al.* Fecal microbiota transplantation from young mice rejuvenates aged hematopoietic stem cells by suppressing inflammation. *Blood* **141**, 1691–1707 (2023).

61. Wu, Z. *et al.* Fecal microbiota transplantation reverses insulin resistance in type 2 diabetes: A randomized, controlled, prospective study. *Front. Cell. Infect. Microbiol.* **12**, 1089991 (2022).

62. Routy, B. *et al.* Fecal microbiota transplantation plus anti-PD-1 immunotherapy in advanced melanoma: a phase I trial. *Nat. Med.* **29**, 2121–2132 (2023).

63. Amorim, N. *et al.* Refining a Protocol for Faecal Microbiota Engraftment in Animal Models After Successful Antibiotic-Induced Gut Decontamination. *Frontiers in Medicine* **9**, (2022).

64. Paramsothy, S. *et al.* Specific Bacteria and Metabolites Associated With Response to Fecal Microbiota Transplantation in Patients With Ulcerative Colitis. *Gastroenterology* **156**, 1440–1454.e2 (2019).

65. Kong, L. *et al.* Linking strain engraftment in fecal microbiota transplantation with maintenance of remission in Crohn's disease. *Gastroenterology* **159**, 2193–2202.e5 (2020).

66. Wang, Y. *et al.* Establishment and resilience of transplanted gut microbiota in aged mice. *iScience* **25**, 103654 (2022).

67. DuPont, H. L. *et al.* Fecal microbiota transplantation in Parkinson's disease-A randomized repeat-dose, placebo-controlled clinical pilot study. *Front. Neurol.* **14**, 1104759 (2023).

68. Bloom, P. P. *et al.* Fecal microbiota transplant improves cognition in hepatic encephalopathy and its effect varies by donor and recipient. *Hepatology Communications* **6**, 2079–2089 (2022).

69. Damman, C. J. *et al.* Low Level Engraftment and Improvement following a Single Colonoscopic Administration of Fecal Microbiota to Patients with Ulcerative Colitis. *PLoS One* **10**, e0133925 (2015).

70. Wilson, B. C. *et al.* Strain engraftment competition and functional augmentation in a multi-donor fecal microbiota transplantation trial for obesity. *Microbiome* **9**, 107 (2021).

71. Doll, J. P. K. *et al.* Fecal Microbiota Transplantation (FMT) as an Adjunctive Therapy for Depression—Case Report. *Front. Psychiatry* **13**, (2022).

72. Ma, J. *et al.* Gut microbiota remodeling improves natural aging-related disorders through Akkermansia muciniphila and its derived acetic acid. *Pharmacol. Res.* **189**, 106687 (2023).

73. Halsey, T. M. *et al.* Microbiome alteration via fecal microbiota transplantation is effective for refractory immune checkpoint inhibitor-induced colitis. *Sci. Transl. Med.* **15**, eabq4006 (2023).

74. Wang, Y. *et al.* Insights into bacterial diversity in compost: Core microbiome and prevalence of potential pathogenic bacteria. *Sci. Total Environ.* **718**, 137304 (2020).

75. Conover, W. J. Practical Nonparametric Statistics, 3rd Edition. https://www.wiley.com/en-us/Practical+Nonparametric+Statistics%2C+3rd+Edition-p-9780471160687.

76. Sørensen, T. J. *A Method of Establishing Groups of Equal Amplitude in Plant Sociology Based on Similarity of Species Content and Its Application to Analyses of the Vegetation on Danish Commons*. (I kommission hos E. Munksgaard, København, 1948).

77. Dagan, I., Lee, L. & Pereira, F. Similarity-based methods for word sense disambiguation. in *Proceedings of the 35th Annual Meeting of the Association for Computational Linguistics and Eighth Conference of the European Chapter of the Association for Computational Linguistics* 56–63 (Association for Computational Linguistics, USA, 1997).

78. Lozupone, C., Lladser, M. E., Knights, D., Stombaugh, J. & Knight, R. UniFrac: an effective distance metric for microbial community comparison. *ISME J.* **5**, 169–172 (2011).

79. Jaccard, P. *Nouvelles Recherches Sur La Distribution Florale*. (Rouge, Lausanne, 1908).

80. Maillet, N., Lemaitre, C., Chikhi, R., Lavenier, D. & Peterlongo, P. Compareads: comparing huge metagenomic experiments. *BMC Bioinformatics* **13**, S10 (2012).

81. Paradis, E. & Schliep, K. ape 5.0: an environment for modern phylogenetics and evolutionary analyses in R. *Bioinformatics* **35**, 526–528 (2019).

82. Bukowski-Thall, G. *et al.* Fecal microbiota transplantation from individual with bipolar disorder and healthy control elicits distinct behaviors and metabolite profiles in mice. *bioRxiv* 2023.11.16.566698 (2023) doi:10.1101/2023.11.16.566698.

83. Freitag, T. L. *et al.* Minor Effect of Antibiotic Pre-treatment on the Engraftment of Donor Microbiota in Fecal Transplantation in Mice. *Front. Microbiol.* **10**, (2019).

84. Tian, D. *et al.* Fecal microbiota transplantation enhances cell therapy in a rat model of hypoganglionosis by SCFA-induced MEK1/2 signaling pathway. *EMBO J.* **42**, e111139 (2023).

85. Wen, X. *et al.* Fecal microbiota transplantation alleviates experimental colitis through the Toll-like receptor 4 signaling pathway. *World J. Gastroenterol.* **29**, 4657–4670 (2023).

86. El-Salhy, M., Patcharatrakul, T. & Gonlachanvit, S. Fecal microbiota transplantation for irritable bowel syndrome: An intervention for the 21st century. *World J. Gastroenterol.* **27**, 2921–2943 (2021).

87. Halko, N., Martinsson, P.-G., Shkolnisky, Y. & Tygert, M. An algorithm for the principal component analysis of large data sets. Preprint at https://doi.org/10.48550/arXiv.1007.5510 (2011).

88. Wang, X.-W. *et al.* Ecological dynamics imposes fundamental challenges in community-based microbial

source tracking. *iMeta* **2**, e75 (2023).

89. Aggarwala, V. *et al.* Precise quantification of bacterial strains after fecal microbiota transplantation delineates long-term engraftment and explains outcomes. *Nature Microbiology* **6**, 1309–1318 (2021).
90. Knights, D. *et al.* Bayesian community-wide culture-independent microbial source tracking. *Nat. Methods* **8**, 761–763 (2011).
91. Liu, J. *et al.* Shifts and importance of viable bacteria in treatment of DSS-induced ulcerative colitis mice with FMT. *Front. Cell. Infect. Microbiol.* **13**, 1124256 (2023).
92. Bryk, A. & Raudenbush, S. Hierarchical linear models: Applications and data analysis methods. Advanced qualitative techniques in the social sciences, 1. **1**, (2002).
93. Hyun, J. *et al.* Faecal microbiota transplantation reduces amounts of antibiotic resistance genes in patients with multidrug-resistant organisms. *Antimicrob. Resist. Infect. Control* **11**, 20 (2022).
94. Mallick, H. *et al.* Multivariable association discovery in population-scale meta-omics studies. *PLoS Comput. Biol.* **17**, e1009442 (2021).
95. Nelder, J. A. & Wedderburn, R. W. M. Generalized Linear Models. *J. R. Stat. Soc. Ser. A* **135**, 370–384 (1972).
96. Aitken, A. C. IV.—On Least Squares and Linear Combination of Observations. *Proceedings of the Royal Society of Edinburgh* **55**, 42–48 (1936).
97. Fisher, R. A. XV.—The Correlation between Relatives on the Supposition of Mendelian Inheritance. (1919) doi:10.1017/s0080456800012163.
98. Wen, X. *et al.* Fecal microbiota transplantation ameliorates experimental colitis via gut microbiota and T-cell modulation. *World J. Gastroenterol.* **27**, 2834–2849 (2021).
99. Stols-Gonçalves, D. *et al.* Faecal Microbiota transplantation affects liver DNA methylation in Non-alcoholic fatty liver disease: a multi-omics approach. *Gut Microbes* **15**, 2223330 (2023).
100. El-Salhy, M., Hatlebakk, J. G., Gilja, O. H., Bråthen Kristoffersen, A. & Hausken, T. Efficacy of faecal microbiota transplantation for patients with irritable bowel syndrome in a randomised, double-blind, placebo-controlled study. *Gut* **69**, 859–867 (2020).

101. Casén, C. *et al.* Deviations in human gut microbiota: a novel diagnostic test for determining dysbiosis in patients with IBS or IBD. *Aliment. Pharmacol. Ther.* **42**, 71–83 (2015).

102. Enck, P. & Mazurak, N. Dysbiosis in Functional Bowel Disorders. *Ann. Nutr. Metab.* **72**, 296–306 (2018).

103. Fernandes, A. D. *et al.* Unifying the analysis of high-throughput sequencing datasets: characterizing RNA-seq, 16S rRNA gene sequencing and selective growth experiments by compositional data analysis. *Microbiome* **2**, 15 (2014).

104. Mann, H. B. & Whitney, D. R. On a Test of Whether one of Two Random Variables is Stochastically Larger than the Other. *Ann. Math. Stat.* **18**, 50–60 (1947).

105. Kang, D.-W. *et al.* Reduced Incidence of Prevotella and Other Fermenters in Intestinal Microflora of Autistic Children. *PLoS One* **8**, e68322 (2013).

106. Adams, J. B., Johansen, L. J., Powell, L. D., Quig, D. & Rubin, R. A. Gastrointestinal flora and gastrointestinal status in children with autism--comparisons to typical children and correlation with autism severity. *BMC Gastroenterol.* **11**, 22 (2011).

107. Finegold, S. M. Desulfovibrio species are potentially important in regressive autism. *Med. Hypotheses* **77**, 270–274 (2011).

108. Costea, P. I. *et al.* Enterotypes in the landscape of gut microbial community composition. *Nature microbiology* **3**, 8–16 (2018).

109. Cheng, M. & Ning, K. Stereotypes About Enterotype: the Old and New Ideas. *Genomics Proteomics Bioinformatics* **17**, 4–12 (2019).

110. Knights, D. *et al.* Rethinking 'Enterotypes'. *Cell Host Microbe* **16**, 433–437 (2014).

111. Yong, E. Gut microbial 'enterotypes' become less clear-cut. *Nature* (2012) doi:10.1038/nature.2012.10276.

112. Garcia, T. P. & Marder, K. Statistical Approaches to Longitudinal Data Analysis in Neurodegenerative Diseases: Huntington's Disease as a Model. *Curr. Neurol. Neurosci. Rep.* **17**, 14 (2017).

113. Bokulich, N. A. *et al.* q2-longitudinal: Longitudinal and Paired-Sample Analyses of Microbiome Data. *mSystems* **3**, (2018).

114. Murphy, J. I., Weaver, N. E. & Hendricks, A. E. Accessible analysis of longitudinal data with linear mixed effects models. *Dis. Model. Mech.* **15**, dmm048025 (2022).

115. Fujimoto, K. *et al.* Functional Restoration of Bacteriomes and Viromes by Fecal Microbiota Transplantation. *Gastroenterology* **160**, 2089–2102.e12 (2021).

116. Podlesny, D. *et al.* Identification of clinical and ecological determinants of strain engraftment after fecal microbiota transplantation using metagenomics. *Cell Rep Med* **3**, 100711 (2022).

117. Schmidt, T. S. B. *et al.* Drivers and determinants of strain dynamics following fecal microbiota transplantation. *Nat. Med.* **28**, 1902–1912 (2022).

118. Ianiro, G. *et al.* Variability of strain engraftment and predictability of microbiome composition after fecal microbiota transplantation across different diseases. *Nat. Med.* **28**, 1913–1923 (2022).

119. Lax, S. *et al.* Bacterial colonization and succession in a newly opened hospital. *Sci. Transl. Med.* **9**, (2017).

# Supplementary Figures

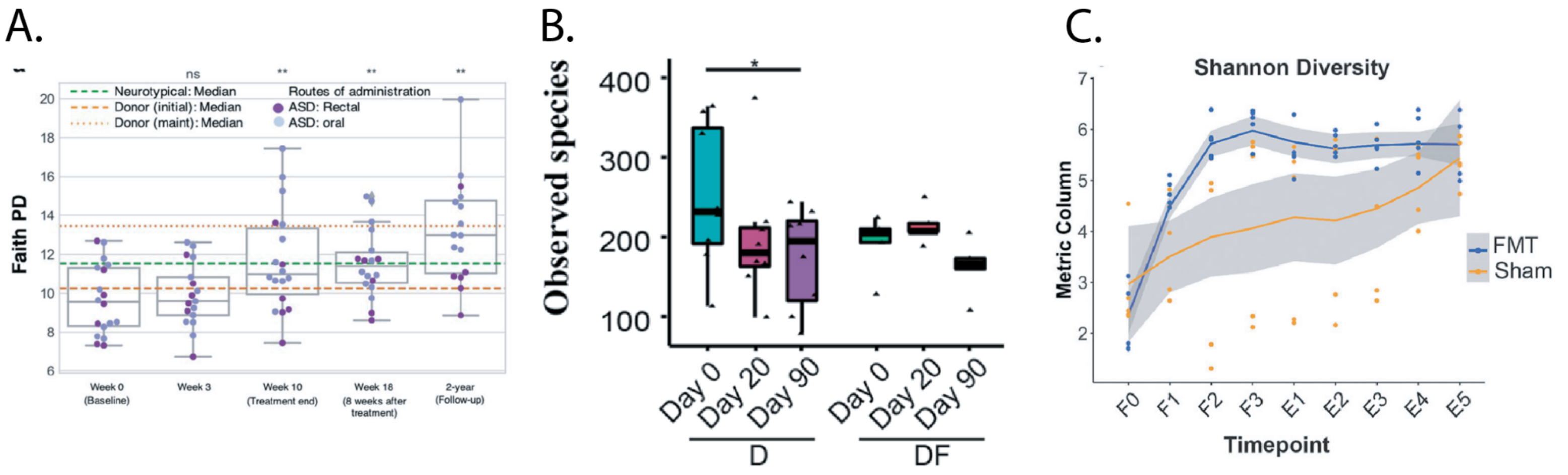

**Supplemental Figure 1: Examples of Common Alpha Diversity Visualizations.** A) Kang et al. 2019 B) Su et al. 2022 C) Amorim et al. 2022. These figures have been reproduced in accordance with their Creative Commons licenses.

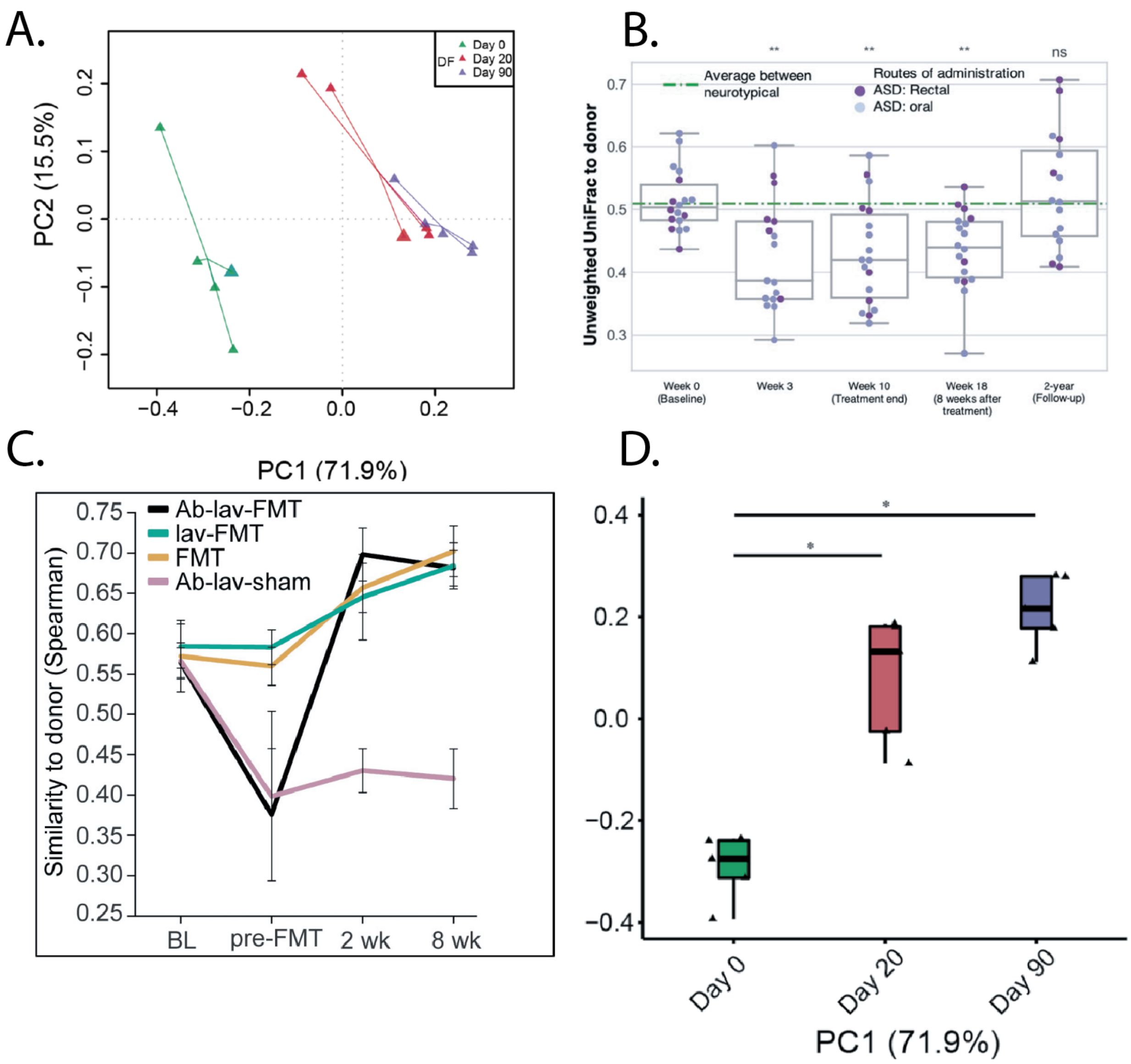

**Supplemental Figure 2: Examples of Common Beta Diversity Visualizations.** A) Su et al. 2022 , B) Kang et al. 2019, C) Freitag et al. 2019, D) Su et al. 2022. These figures have been reproduced in accordance with their Creative Commons licenses.

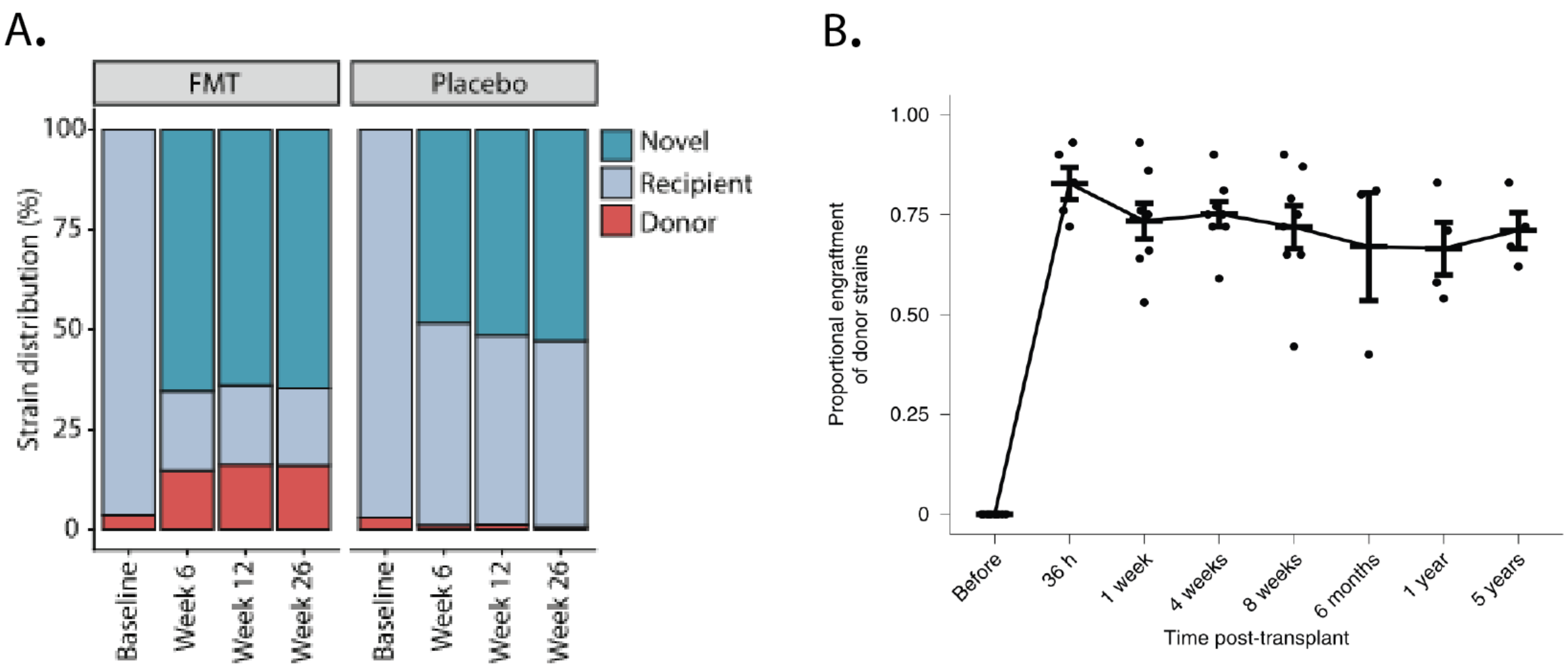

**Supplemental Figure 3: Examples of Common FMT Source Tracking Visualizations.** A) Wilson et al. 2021, B) Aggarwala et al. 2021. These figures have been reproduced in accordance with their Creative Commons licenses.

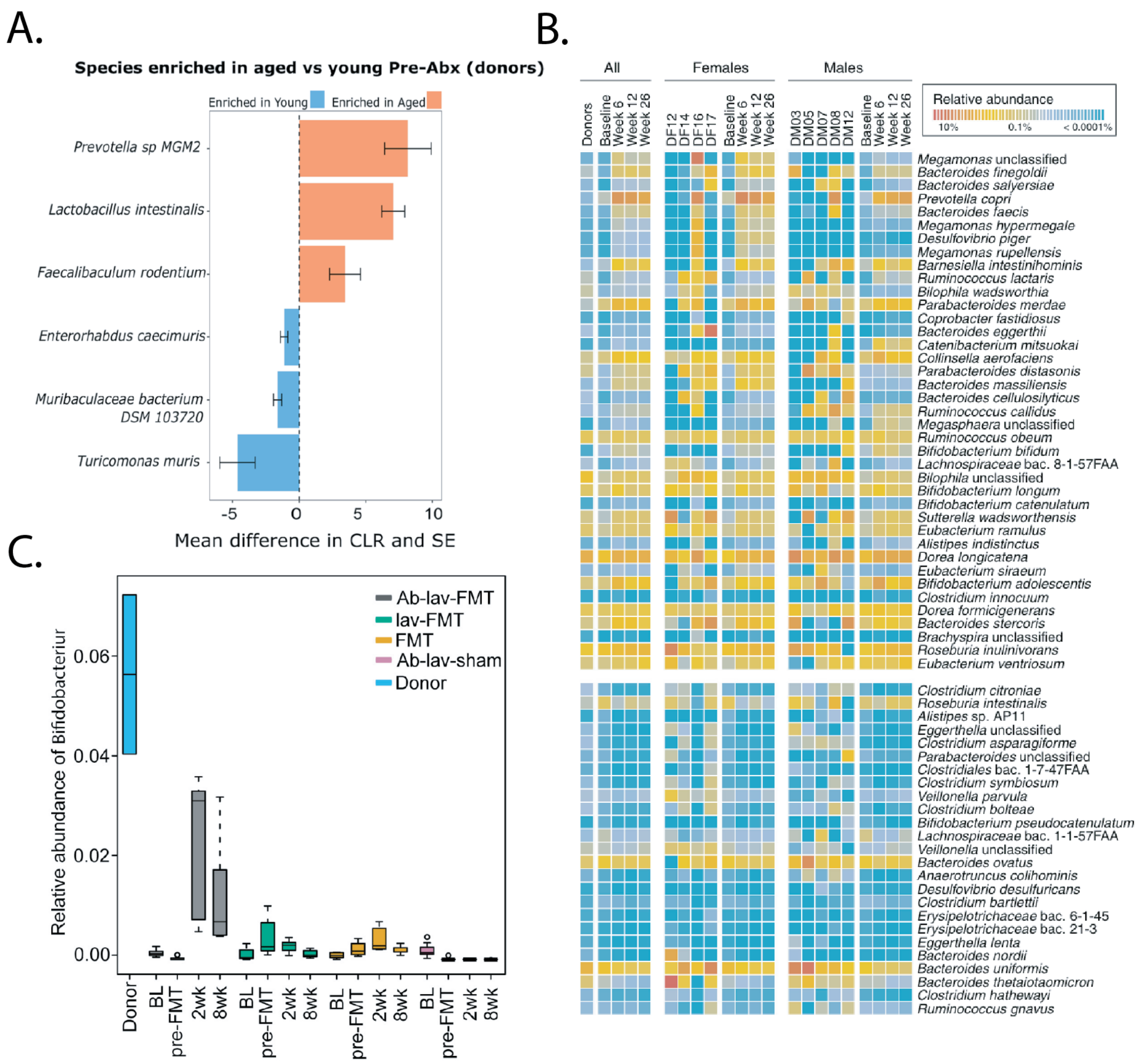

**Supplemental Figure 4: Examples of Common Feature Abundance Visualizations.** A) Parker et al. 2022, B) Wilson et al. 2021, C) Freitag et al. 2019. These figures have been reproduced in accordance with their Creative Commons licenses.

# Supplemental Data A

| | | |
|---|---|---|
| Alpha Diversity | Li et al., 2018<br>Singh et al., 2022<br>Gopalakrishnan et al., 2021<br>Parker et al., 2022<br>Bloom et al., 2021<br>Baruch et al., 2020<br>Doll et al., 2022<br>Su et al., 2022<br>Kang et al., 2019<br>Kang et al., 2017<br>Freitag et al., 2019<br>Aggarwala et al., 2021<br>Amorim et al., 2022<br>Wilson et al., 2021<br>Hazan et al., 2021 | DeFilipp et al., 2018<br>Damman et al., 2015<br>Wang et al., 2022<br>Park et al., 2021<br>Riquelme et al., 2019<br>Lier et al., 2020<br>Staley et al., 2019<br>Zeng et al., 2023<br>Liu et al., 2023<br>Wu et al., 2022<br>Routy et al., 2023<br>Ma et al., 2023<br>Kong et al., 2020<br>Yu et al ., 2023<br>Halsey et al., 2023<br>Bukowski-Thall et al ., 2023<br>Dupont et al., 2023<br>Paramsothy et al., 2019 |
| Beta Diversity | Li et al., 2018<br>Singh et al., 2022<br>Gopalakrishnan et al., 2021<br>Parker et al., 2022<br>Bloom et al., 2021<br>Davar et al., 2021<br>Baruch et al., 2020<br>El-Salhy et al., 2021<br>Su et al., 2022<br>Kang et al., 2019<br>El-Salhy et al., 2020<br>Kang et al., 2017<br>Freitag et al., 2019<br>Amorim et al., 2022<br>Wilson et al., 2021<br>Hazan et al., 2021<br>DeFilipp et al., 2018 | Damman et al., 2015<br>Wang et al., 2022<br>Park et al., 2021<br>Riquelme et al., 2019<br>Schmidt et al., 2022<br>Staley et al., 2019<br>El-Salhy et al., 2022<br>Zeng et al., 2023<br>Liu et al., 2023<br>Wu et al., 2022<br>Routy et al., 2023<br>Ma et al., 2023<br>Kong et al., 2020<br>Halsey et al., 2023<br>Bukowski-Thall et al., 2023<br>Dupont et al., 2023<br>Tian et al ., 2022<br>Wen et al., 2023<br>Paramsothy et al ., 2019 |
| Differential Abundance | Li et al., 2018<br>Parker et al., 2022<br>Bloom et al., 2021<br>Davar et al., 2021<br>Baruch et al., 2020<br>Doll et al., 2022<br>Su et al., 2022<br>El-Salhy et al., 2020<br>Freitag et al., 2019<br>Wilson et al., 2021<br>Wang et al., 2022<br>Park et al., 2021<br>Riquelme et al., 2019<br>El-Salhy et al., 2022 | Zeng et al., 2023<br>Liu et al., 2023<br>Wu et al., 2022<br>Routy et al., 2023<br>Ma et al., 2023<br>Kong et al., 2020<br>El-Salhy et al., 2021<br>Halsey et al., 2023<br>Bukowski-Thall et al., 2023<br>Dupont et al., 2023<br>Tian et al., 2022<br>Wen et al ., 2023<br>Wen et al., 2021<br>Paramsothy et al., 2019<br>Hyun et al., 2022 |
| Literature Identified Features | Kang et al., 2019<br>Kang et al., 2017<br>Wilson et al., 2021<br>DeFilipp et al., 2018<br>Lier et al., 2020 | |
| Source Tracking | Singh et al., 2022<br>Gopalakrishnan et al., 2021<br>Aggarwala et al., 2021<br>Wilson et al., 2021<br>DeFilipp et al., 2018<br>Riquelme et al., 2019<br>Staley et al., 2019<br>Routy et al., 2023 | |

# Supplemental Data B

| Title | Study Length (days) |
| --- | --- |
| Clinical Efficacy and Microbiome Changes Following Fecal Microbiota Transplantation in Children With Recurrent Clostridium Difficile Infection | 30 |
| Effect of antibiotic pretreatment on bacterial engraftment after Fecal Microbiota Transplant (FMT) in IBS-D | 70 |
| Engraftment of Bacteria after Fecal Microbiota Transplantation Is Dependent on Both Frequency of Dosing and Duration of Preparative Antibiotic Regimen | 21 |
| Fecal microbiota transfer between young and aged mice reverses hallmarks of the aging gut, eye, and brain | 18 |
| Fecal microbiota transplant improves cognition in hepatic encephalopathy and its effect varies by donor and recipient | 45 |
| Fecal microbiota transplant overcomes resistance to anti–PD-1 therapy in melanoma patients | 560 |
| Fecal microbiota transplant promotes response in immunotherapy-refractory melanoma patients | 65 |
| Fecal Microbiota Transplantation (FMT) as an Adjunctive Therapy for Depression—Case Report | 56 |
| Fecal microbiota transplantation for irritable bowel syndrome: An intervention for the 21st century | 30 |
| Health improvements of type 2 diabetic patients through diet and diet plus fecal microbiota transplantation | 90 |
| Long-term benefit of Microbiota Transfer Therapy on autism symptoms and gut microbiota | 712 |
| Long-term effects of fecal microbiota transplantation (FMT) in patients with irritable bowel syndrome | 356 |
| Microbiota Transfer Therapy alters gut ecosystem and improves gastrointestinal and autism symptoms: an open-label study | 126 |
| Minor Effect of Antibiotic Pre-treatment on the Engraftment of Donor Microbiota in Fecal Transplantation in Mice | 56 |
| Precise quantification of bacterial strains after fecal microbiota transplantation delineates long-term engraftment and explains outcomes | 1825 |
| Refining a Protocol for Faecal Microbiota Engraftment in Animal Models After Successful Antibiotic-Induced Gut Decontamination | 56 |
| Strain engraftment competition and functional augmentation in a multi-donor fecal microbiota transplantation trial for obesity | 182 |
| Successful Bacterial Engraftment Identified by NextGeneration Sequencing Predicts Success of Fecal Microbiota Transplant for Clostridioides difficile | 30 |
| Third-party fecal microbiota transplantation following allo-HCT reconstitutes microbiome diversity | 356 |
| Low Level Engraftment and Improvement following a Single Colonoscopic Administration of Fecal Microbiota to Patients with Ulcerative Colitis | 30 |

Establishment and resilience of transplanted gut microbiota in aged mice

56

Cognitive function improvement after fecal microbiota transplar

21

Tumor Microbiome Diversity and Composition Influence Pancreatic Cancer Outcomes

35

Donor fecal microbiota transplantation ameliorates intestinal graft-versus-host disease in allogeneic hematopoietic cell transplant recipients

168

Drivers and determinants of strain dynamics following fecal microbiota transplantation

159.4

Durable Long-Term Bacterial Engraftment following Encapsula

409

Efficacy of faecal microbiota transplantation for patients with ir

84

Linking strain engraftment in fecal microbiota transplantation with maintenance of remission in Crohn’s disease

168

Reconstitution of the gut microbiota of antibiotic-treated patients by autologous fecal microbiota transplant

50

Fecal microbiota transplantation from young mice rejuvenates aged hematopoietic stem cells by suppressing inflammation

21

Shifts and importance of viable bacteria in treatment of DSS-induced ulcerative colitis mice with FMT

15

| | |
|---|---|
| Fecal microbiota transplantation reverses insulin resistance in type 2 diabetes: A randomized, controlled, prospective study | 30 |
| Fecal microbiota transplantation plus anti-PD-1 immunotherapy in advanced melanoma: a phase I trial | 120 |
| Gut microbiota remodeling improves natural aging-related disorders through Akkermansia muciniphila and its derived acetic acid | 84 |
| Successful Bacterial Engraftment Identified by Next-Generation Sequencing Predicts Success of Fecal Microbiota Transplant for Clostridioides difficile | 30 |
| Microbiota Transfer Therapy alters gut ecosystem and improves gastrointestinal and autism symptoms: an open-label study | 112 |
| Fecal microbiota transplantation inhibits colorectal cancer progression: Reversing intestinal microbial dysbiosis to enhance anti-cancer immune responses | 7 |
| Microbiome alteration via fecal microbiota transplantation is effective for refractory immune checkpoint inhibitor–induced colitis | 28 |
| Fecal microbiota transplantation in Parkinson's disease—A randomized repeat-dose, placebo-controlled clinical pilot study | 365 |
| Specific Bacteria and Metabolites Associated With Response to Fecal Microbiota Transplantation in Patients With Ulcerative Colitis | 56 |
| Faecal microbiota transplantation reduces amounts of antibiotic resistance genes in patients with multidrug-resistant organisms | 28 |